\documentclass[conference]{IEEEtran}
\IEEEoverridecommandlockouts
\def\BibTeX{{\rm B\kern-.05em{\sc i\kern-.025em b}\kern-.08em
    T\kern-.1667em\lower.7ex\hbox{E}\kern-.125emX}}
\usepackage{cite}
\usepackage{amsmath,amssymb,amsfonts}
\usepackage{graphicx}
\usepackage{subcaption}
\usepackage[table]{xcolor}
\usepackage{booktabs}
\usepackage{placeins}
\usepackage{enumitem}
\begin{document}

\title{PLRTune: Importance Pre-Sampling and LLM-Guided Reinforcement Learning for Automatic Database Tuning%
\thanks{*Chen Zheng is the corresponding author.}}

\author{\IEEEauthorblockN{
Xinyue Yang\textsuperscript{1,2},
Chen Zheng\textsuperscript{1,2,3,*},
Yaoyang Hou\textsuperscript{3},
Renhao Zhang\textsuperscript{3},\\
Yinyan Zhang\textsuperscript{2},
Heng Zhang\textsuperscript{2}}
\IEEEauthorblockA{
\textsuperscript{1}\textit{University of Chinese Academy of Sciences, Nanjing, China}\\
\textsuperscript{2}\textit{Institute of Software, Chinese Academy of Sciences, Beijing, China}\\
\textsuperscript{3}\textit{Hangzhou Institute for Advanced Study, UCAS, Hangzhou, China}
}
}

\maketitle



\begin{abstract}

Configuration tuning is critical to database performance, yet automatic database tuning remains challenging due to high-dimensional knob spaces, substantial online tuning cost, unreliable textual hints derived from Large Language Models (LLMs) or community documents, and the difficulty of exploiting the remaining optimization room after initialization.

Hence, we propose PLRTune, a staged database tuning system that leverages workload-specific domain knowledge to identify a reduced search space and further optimize within this promising region. First, we develop an importance pre-sampling and reranking strategy to identify the dominant knob subset in a workload-specific manner and derive a compact state representation. Second, we design an execution-guided hint refinement technique to improve the initialization quality of documentation-guided tuning. Finally, we propose a post-tuning refinement stage that leverages Twin Delayed Deep Deterministic Policy Gradient (TD3) to explore the dominant knob subset and further exploit the remaining optimization room.

We evaluate PLRTune on MySQL and PostgreSQL across diverse benchmark workloads. Compared with state-of-the-art approaches, PLRTune achieves the best final result on all tested workloads, improving over the corresponding best-performing alternative by \textbf{9.50\%} on average. Moreover, PLRTune reaches the strongest baseline's best performance level \textbf{9.03$\times$ faster} on average across workloads, demonstrating its practical runtime efficiency without sacrificing final tuning quality.

\end{abstract}



\section{Introduction}
\label{sec:introduction}

Modern database management systems (DBMSs) underpin a wide range of data-intensive applications \cite{zhao2023automatic}. Their performance is highly sensitive to configuration parameters, commonly referred to as knobs \cite{zhang2022facilitating,kanellis2020too}. Systems such as MySQL and PostgreSQL expose hundreds of tunable knobs controlling memory, execution, logging, and concurrency behavior. Because these knobs are high-dimensional, strongly interdependent, and sensitive to workload and hardware characteristics, manual tuning by database administrators (DBAs) is time-consuming and often yields suboptimal configurations \cite{duan2009tuning,CDBTune,storm2006adaptive}.

To reduce the burden of manual tuning, prior work has explored automatic database tuning from multiple directions \cite{cgptuner,duan2009tuning,gur2021adaptive,li2019qtune,zhao2023automatic,wang2021udo,BestConfig,zhang2021restune}. Broadly, existing studies can be understood from three perspectives. The first perspective is \emph{data-driven tuning}, where systems such as OtterTune \cite{van2017automatic} and E2ETune \cite{huang2024e2etune} perform prior data collection and model learning to build reusable tuning knowledge from historical observations. The second perspective is \emph{iterative optimization-based tuning}, where methods directly tune a target workload by repeatedly selecting configurations, running workloads, and updating the optimizer according to observed performance. These methods typically formulate tuning as an online black-box optimization or sequential decision-making problem, and progressively improve configurations by balancing exploration of unseen regions with exploitation of accumulated feedback. Typical examples include Bayesian Optimization methods \cite{duan2009tuning,zhang2022facilitating,wang2021udo} and Reinforcement Learning methods \cite{CDBTune,HUNTER,gur2021adaptive,li2019qtune}. The third perspective is \emph{documentation-driven} or \emph{knowledge-guided tuning}, where methods such as DB-BERT and GPTuner \cite{DB-BERT,GPTuner} leverage textual resources including DBMS manuals, engineering blogs, and community discussions to extract tuning hints and guide knob selection and value recommendation. Together, these studies show that automatic database tuning can benefit from offline knowledge reuse, online optimization, and external domain knowledge.

Despite recent progress, efficient automatic database tuning still faces several practical challenges.

\textbf{C1. Low tuning efficiency due to high-dimensional knob spaces.}
Modern DBMSs expose hundreds of tunable knobs, which collectively define a vast and high-dimensional configuration space. Searching such a space can significantly reduce tuning efficiency by slowing down the optimization process and increasing tuning cost. Yet, for a specific workload, only a small subset of knobs often has dominant impact on performance, while many others contribute little \cite{van2017automatic,kanellis2022llamatune}. This makes it important to focus tuning on workload-relevant knobs rather than treating all exposed knobs uniformly from the beginning.

\textbf{C2. Limited starting quality of existing warm-start strategies.}
Many online tuning methods suffer from cold start and may require numerous ineffective trials before reaching competitive configurations \cite{CDBTune,HUNTER,li2019qtune}. Although some methods introduce warm-start mechanisms, such as GA-based sample generation, such initialization may still be insufficient under strict online tuning budgets, especially when tuning is conducted in a restart-based single-server setting without extensive cloned or parallel exploration \cite{HUNTER}. Consequently, the starting quality of tuning remains an important factor that affects both convergence speed and practical tuning cost.

\textbf{C3. Low-quality hints in text-guided tuning.}
Documentation-guided methods rely heavily on the quality of textual hints~\cite{DB-BERT,GPTuner}. In practice, however, not all extracted hints are useful for a specific workload or DBMS setting. To examine this issue, we conduct a small motivating analysis using the raw DB-BERT-style hint set before execution-guided refinement. For each DBMS, we run the analysis on three representative Sysbench workloads, namely read-only, read-write, and write-only, and average the observed parameter impact across these workloads. For each hinted parameter, we analyze whether it is selected during initialization and whether its average observed impact is positive or negative. A parameter is marked as \emph{Negative} when its selected trials show lower average objective values than unselected trials, and as \emph{Never Used} when it is never selected during initialization.

\begin{table}[t]
\centering
\caption{Raw DB-BERT-style hint quality averaged over three Sysbench workloads.}
\label{tab:motivating-raw-hint-impact}
\scriptsize
\setlength{\tabcolsep}{5pt}
\begin{tabular}{lcccc}
\toprule
\textbf{DBMS} &
\shortstack{\textbf{Raw Hinted}\\\textbf{Params}} &
\shortstack{\textbf{Negative}\\\textbf{Params}} &
\shortstack{\textbf{Never}\\\textbf{Used}} &
\shortstack{\textbf{Neg. or}\\\textbf{Unused}} \\
\midrule
MySQL & 30 & 17 & 4 & 21 (70.0\%) \\
PostgreSQL & 24 & 11 & 2 & 13 (54.2\%) \\
\bottomrule
\end{tabular}
\end{table}

As shown in Table~\ref{tab:motivating-raw-hint-impact}, a large fraction of raw hinted parameters are either associated with negative impact or never selected in the initialization traces. Specifically, this happens for 70.0\% of the raw hinted parameters in MySQL and 54.2\% in PostgreSQL. This result suggests that textual hints can provide useful prior knowledge, but they should not be blindly treated as fixed tuning rules. Directly following low-value or context-mismatched hints may mislead the tuning process, resulting in poor initialization and slower tuning progress.

\textbf{C4. Remaining optimization room after knowledge-guided tuning.}
Documentation-guided methods can leverage textual hints to achieve rapid performance gains at the early stage of tuning \cite{DB-BERT,GPTuner}. However, even with a strong initial boost, the configurations derived from textual guidance are often still suboptimal. This is because text-guided tuning is naturally limited by the hinted knobs, recommended value regions, and prior knowledge extracted from external documents or language models. It does not systematically exploit the remaining continuous configuration space through execution feedback. Thus, a significant portion of the achievable performance may remain beyond what knowledge-guided initialization alone can close.

To address these challenges, we propose PLRTune, a three-phase database tuning system that follows a staged warmup-to-fine-tuning pipeline. The system progressively moves through three phases: importance pre-sampling and reranking, execution-guided hint refinement, and post-tuning refinement. In this way, PLRTune reduces search-space complexity, improves early-stage initialization quality, and further exploits the remaining optimization room within a unified pipeline.

\textbf{Phase 0: Importance Pre-Sampling and Reranking.} To address \textbf{C1}, we propose an importance pre-sampling and reranking phase that first uses Latin hypercube sampling (LHS) to obtain well-distributed samples from the validated candidate knob space and collect diverse workload observations into a shared pool. PLRTune then applies Random Forest to score and rerank knobs according to their workload relevance, and further uses PCA to derive a compact state representation from internal metrics. This phase also identifies recurring high-impact knobs across workloads as a reusable workload-aware prior. In this way, Phase~0 identifies performance-critical knobs, reduces the effective search space, and mitigates the tuning inefficiency caused by high-dimensional knob spaces.

\textbf{Phase 1: Execution-Guided Hint Refinement.} To address \textbf{C2} and \textbf{C3}, we propose an execution-guided hint-refinement phase that refines textual hints through execution-guided filtering and empirical recommendations. This phase improves initialization quality, enabling PLRTune to start from a stronger performance level while reducing the risk that low-quality hints mislead the tuning process and delay convergence.

\textbf{Phase 2: Post-Tuning Refinement.} To address \textbf{C4}, we perform post-tuning refinement within the reduced knob subspace identified in Phase~0, starting from the stronger region obtained after Phase~1. Specifically, PLRTune applies TD3 as a state-aware refinement optimizer in this reduced space to further exploit the remaining tuning room and improve final tuning quality after the initial performance gains have already been achieved.

We conduct extensive experiments on MySQL and PostgreSQL across nine workloads, including TPC-H, TPC-C, YCSB, and Sysbench variants. The evaluation covers end-to-end tuning quality, convergence behavior, ablation studies, runtime efficiency, and cross-hardware reuse. PLRTune achieves the best final performance on all tested workloads, improving over the corresponding best baselines by \textbf{9.50\%} on average. Moreover, PLRTune reaches the strongest baseline's best performance level \textbf{9.03$\times$ faster} on average across workloads, while still achieving better final performance on every workload. It also remains effective on both lower-resource and higher-resource target servers.

The key contributions of this work are summarized as follows:
\begin{enumerate}[leftmargin=*, labelsep=0.5em, itemsep=0pt, topsep=2pt, parsep=0pt, partopsep=0pt]



\item \textbf{We propose PLRTune, a novel staged database tuning system that leverages workload-specific domain knowledge to improve database tuning.}
PLRTune provides a unified tuning pipeline that coordinates workload-specific knowledge and execution feedback across different tuning stages. The system supports the complete transition from identifying promising tuning directions, to constructing strong initial configurations, and to further improving configuration quality under a limited tuning budget.

\item \textbf{We design a workload-aware tuning strategy that integrates prior construction, execution-guided initialization, and reduced-space refinement.}
PLRTune constructs workload-specific knob and state priors through importance pre-sampling, Random-Forest-based knob reranking, and PCA-based state compression, reducing unnecessary exploration in the original high-dimensional space. It further treats documentation-derived hints as candidates to be validated rather than fixed rules: execution feedback is used to adapt hint values, prune low-value hints, and strengthen initialization with empirical priors. Starting from this stronger initialization, TD3-based post-tuning refinement exploits the residual optimization room within the reduced knob subspace.

\item \textbf{We conduct extensive experiments and cost analysis to evaluate the effectiveness and portability of PLRTune.}
We evaluate PLRTune on MySQL and PostgreSQL across diverse benchmark workloads and hardware settings. Compared with state-of-the-art methods, PLRTune achieves the best final tuning performance on all tested workloads while reaching strong configurations within a shorter tuning time.

\end{enumerate}

\section{Motivation}
\label{sec:motivation}

We further motivate the design of PLRTune from three observations. These observations explain why efficient automatic database tuning requires workload-aware search-space reduction, strong early-stage initialization, and post-tuning refinement after text-guided tuning.

\textbf{M1: Workload-aware knob identification and dimensionality reduction are critical for efficient tuning.}
DBMS performance is typically dominated by a small subset of knobs, and this subset can vary substantially across workloads. Prior work has observed that knob importance should be treated in a workload-aware manner rather than reused uniformly across workloads~\cite{kanellis2022llamatune}. Directly tuning the full knob space is therefore inefficient: the search cost grows with dimensionality, while the performance gain often diminishes quickly as additional low-impact knobs are included~\cite{van2017automatic,kanellis2022llamatune}. HUNTER addresses this issue by applying Random Forest for knob ranking and principal component analysis (PCA) for metric compression, showing that dimensionality reduction can effectively reduce tuning overhead~\cite{HUNTER}. Moreover, compared with reusing a concrete final configuration, which is tightly coupled to specific hardware, data, and runtime states, reusing workload-relevant important knobs provides a more suitable prior for later tuning. Motivated by these observations, our first phase collects workload-specific observations, applies Random-Forest-based analysis to identify important knobs, uses PCA to derive a compact state representation, and retains the resulting knob subset as a reusable workload-aware prior for subsequent tuning.

\textbf{M2: Strong early-stage initialization should focus on the most influential knobs.}
Prior work suggests that tuning is more effective when it focuses on high-impact knobs rather than treating all knobs equally from the beginning~\cite{van2017automatic,HUNTER}. For example, OtterTune adopts an incremental tuning strategy and shows that it performs better than directly tuning either too few or too many knobs at once~\cite{van2017automatic}. In addition, DB-BERT uses textual hints to guide a small number of important knobs toward recommended directions, enabling rapid early convergence~\cite{DB-BERT}. These observations motivate our second phase: we use text-guided tuning as an intermediate initialization stage, so that the system can quickly move workload-relevant knobs toward promising regions and alleviate the low efficiency caused by cold start.

\textbf{M3: Text-guided tuning still leaves residual optimization room.}
Prior text-guided methods show that database performance can be improved quickly by adjusting a small number of important knobs according to textual hints~\cite{DB-BERT,GPTuner}. This indicates that focusing on hinted high-impact knobs is an effective way to obtain rapid gains. However, such guidance is naturally limited by the hinted knobs and recommended value regions. GPTuner also suggests that methods such as DB-BERT mainly adjust a limited set of hinted knobs, and therefore may leave substantial optimization space unexplored~\cite{GPTuner}. To further validate this issue in our setting, we conduct a small motivating experiment on MySQL Sysbench Read-Write: after running DB-BERT for 8 initialization steps, we use the resulting configuration as the starting point and continue tuning with TD3-based post-tuning refinement.

\begin{figure}[t]
\centering
\includegraphics[width=0.60\columnwidth]{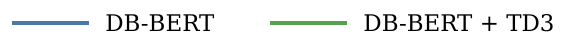}
\includegraphics[width=0.65\columnwidth]{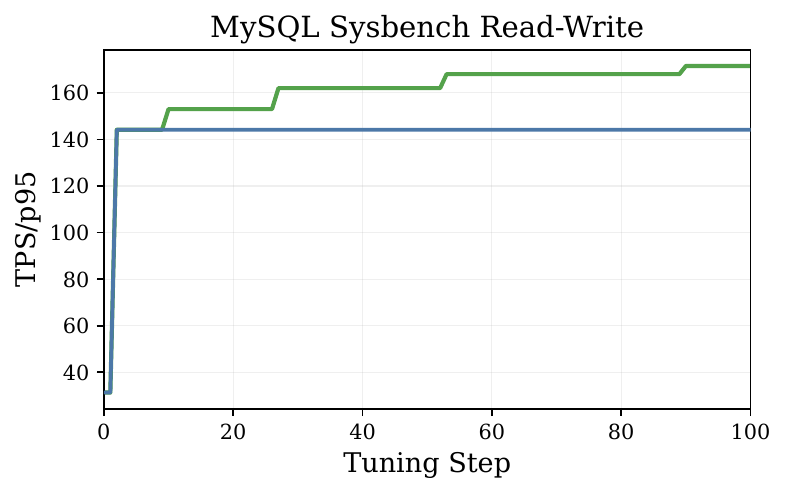}
\captionsetup{skip=2pt}
\caption{Motivating experiment comparing DB-BERT initialization with TD3-based post-tuning on MySQL Sysbench Read-Write.}
\label{fig:motivation-post-tuning}
\end{figure}

As shown in Fig.~\ref{fig:motivation-post-tuning}, DB-BERT quickly reaches a strong best-so-far value and then remains nearly flat, while DB-BERT+TD3 continues to improve the objective value from about 144 to around 170. This result shows that the configuration produced by text-guided tuning is a strong starting point, but it does not fully close the remaining optimization space. Therefore, after text-guided initialization, tuning should continue to exploit residual opportunities through execution-guided refinement.

\section{Preliminaries}
\label{sec:preliminaries}

Consistent with prior database tuning systems, we represent each tuning trial using a unified tuple \((S,A,P)\), where \(A\) denotes the action taken in the trial, i.e., the database configuration applied to the DBMS, \(S\) denotes the internal state observed during the trial, and \(P\) denotes the resulting external performance.

\paragraph{\textbf{A (Action / Configuration)}}
\(A\) represents the vector of tunable database knobs applied in one trial. In this work, we first construct a practical candidate action space from the exposed DBMS variables through lightweight LLM-assisted screening followed by deterministic validation. The validation step removes knobs that are unsupported in the target DBMS version, unsafe to modify, have invalid value domains, or are incompatible with the single-node evaluation protocol used in our experiments. After this construction step, we use 150 knobs for MySQL~8.0.45 and 100 knobs for PostgreSQL~16.13 as the practical candidate action spaces throughout the study. These spaces cover major performance-relevant categories, including memory management, storage-engine behavior, checkpoint/WAL or logging controls, connection and thread management, cache-related settings, optimizer-related parameters, and selected monitoring or timeout controls. This follows prior tuning practice that constructs filtered candidate spaces before further dimensionality reduction~\cite{kanellis2022llamatune}.

\paragraph{\textbf{P (Performance Metrics)}}
\(P\) denotes the external performance of the database under configuration \(A\). In our experiments, we measure throughput and tail latency, including TPS, QPS, and p95 latency, for transactional and mixed workloads, while using execution time for TPC-H. The specific scalar objective used for optimization is defined later in the experimental setup.


\paragraph{\textbf{S (State / Internal Metrics)}}
\(S\) represents the internal state of the DBMS collected during one tuning trial. For MySQL, we query enabled InnoDB metrics from \texttt{information\_schema.INNODB\_METRICS}, resulting in 74 internal metrics in MySQL~8.0.45. For PostgreSQL~16.13, we collect 112 internal metrics from system statistics views, including \texttt{pg\_stat\_database}, \texttt{pg\_stat\_bgwriter}, \texttt{pg\_stat\_activity}, and related WAL, I/O, lock, and conflict statistics views.

To reduce measurement noise, internal metrics are collected over multiple frames during each benchmark run. Let \(m_i^t\) denote the value of metric \(m_i\) at frame \(t \in \{1,2,\ldots,T\}\). We aggregate the temporal observations into a single scalar \(s_i\) for each metric according to whether the metric is cumulative or instantaneous.

For cumulative counters, which monotonically increase over time, we use the difference between the last and first frames:
\begin{equation}
s_i^{\mathrm{counter}} = m_i^T - m_i^1.
\end{equation}

For instantaneous or gauge-style metrics, which reflect the current operating state and may fluctuate during a run, we use the temporal average:
\begin{equation}
s_i^{\mathrm{gauge}} = \frac{1}{T}\sum_{t=1}^{T} m_i^t.
\end{equation}

The resulting state vector is then written as
\[
S = [s_1, s_2, \ldots, s_d],
\]






where \(d=74\) for MySQL and \(d=112\) for PostgreSQL before dimensionality reduction. In PLRTune, this state representation is further compressed by PCA to improve the efficiency and stability of the later post-tuning refinement stage.

\paragraph{\textbf{Shared Pool}}
All tuples \((S, A, P)\) generated during Phase~0 importance pre-sampling and subsequent tuning are stored in a shared pool for later reuse across different stages of the system.

\section{System Architecture and Workflow}
\label{sec:architecture}

PLRTune is designed as a staged tuning pipeline that separates prior construction, initialization, and final refinement. Given a target DBMS and workload, PLRTune repeatedly applies candidate configurations, executes the workload, collects internal state metrics and external performance, and stores the resulting tuples \((S, A, P)\) in a Shared Pool. The key design idea is to avoid optimizing the full knob space from scratch. Instead, PLRTune first constructs a workload-specific reduced action and state space, then uses execution-guided textual hints to obtain a strong initialization, and finally performs state-aware reinforcement-learning refinement within the reduced subspace.

For a newly arriving workload, PLRTune may optionally use lightweight workload-type matching to retrieve a relevant knob prior from the Shared Pool. This step compares simple workload-composition features with historical workloads and selects the closest workload type. The matched result is used only to identify or reuse a candidate set of workload-relevant knobs, rather than to transfer concrete knob values or a final configuration. The actual tuned values are still determined through execution feedback in the following phases.

\begin{figure*}[t]
\centering
\includegraphics[width=0.7\textwidth]{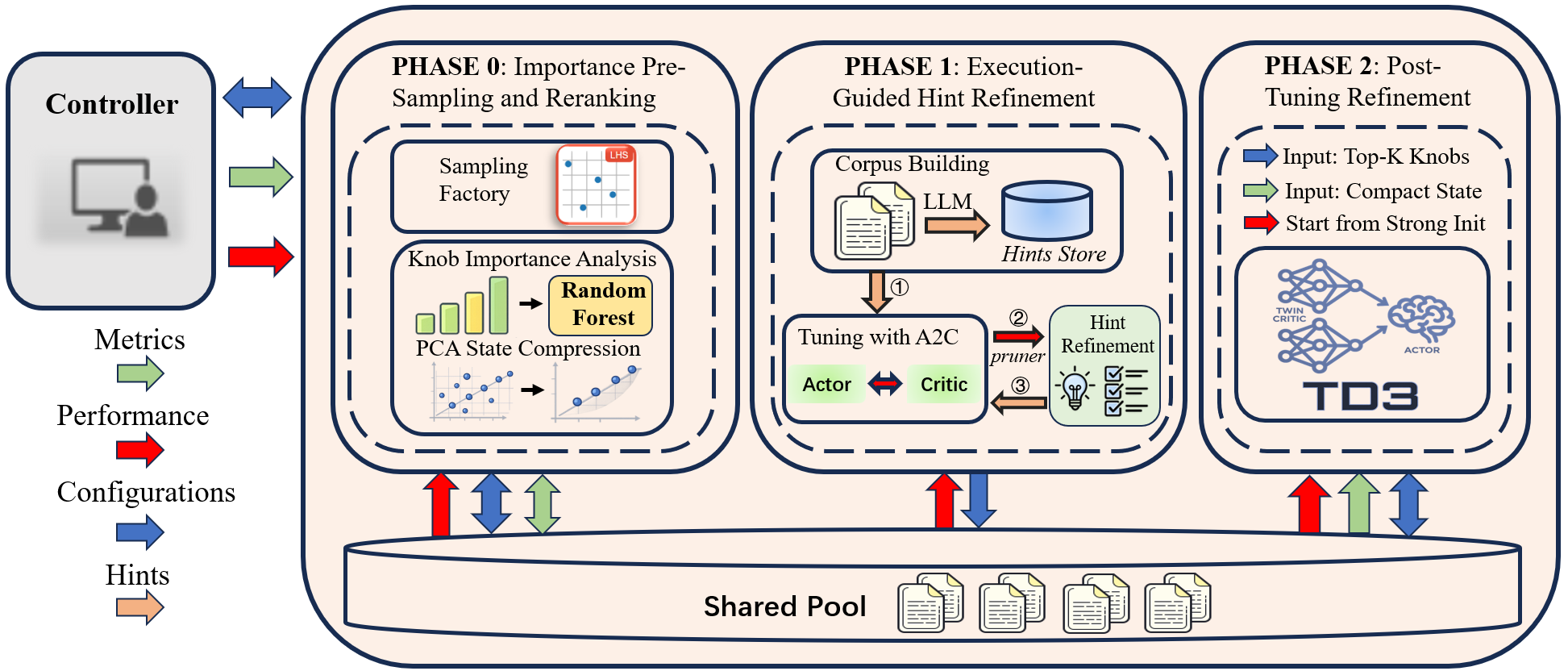}
\captionsetup{skip=2pt}
\caption{System architecture and workflow of PLRTune.}
\label{fig:framework}
\end{figure*}

Figure~\ref{fig:framework} illustrates the overall architecture of PLRTune. The system is centered around the Shared Pool, which stores tuning tuples and connects three stages: Phase~0 constructs workload-specific knob and state priors, Phase~1 validates and refines documentation-derived hints to build a stronger initialization, and Phase~2 performs state-aware refinement in the reduced subspace. Together, these stages use workload-specific priors and execution feedback to reduce unnecessary search and exploit the remaining optimization room.

\subsection{Phase 0: Importance Pre-Sampling and Reranking}
\label{sec:phase0}

The goal of Phase~0 is to construct a workload-specific reduced tuning subspace before the subsequent stages of PLRTune. Rather than relying on the online optimizer to discover important knobs and compact state representations from scratch, PLRTune first collects diverse workload observations over the validated candidate knob space and uses them to build workload-aware priors. This phase serves two purposes: (1) it provides a broad and well-distributed sample basis for estimating knob relevance, and (2) it derives reduced action and state representations that make later online refinement more focused and efficient.

For each DBMS version and workload type, PLRTune generates candidate configurations using LHS. Compared with naive random sampling, LHS encourages more uniform coverage of the high-dimensional candidate knob space, which reduces the risk that the subsequent importance analysis is biased by a narrow set of early configurations. In each step, a sampled knob vector is mapped to an executable DBMS configuration, applied to the target system, and evaluated under the corresponding benchmark workload. During execution, PLRTune records the applied knob values, the aggregated internal state metrics, and the resulting external performance into the Shared Pool in the unified tuple form \((S,A,P)\).

Based on the collected samples, Phase~0 reduces the later tuning burden through two concrete analyses. First, PLRTune trains a Random-Forest-based performance model on the pre-sampled observations and uses the resulting feature-importance scores to rerank knobs according to workload relevance. The Top-$K$ knobs with the largest importance scores are retained as the reduced action subspace for later stages. This design allows subsequent optimization to focus on influential knobs instead of treating all candidate knobs uniformly. Second, PLRTune applies principal component analysis (PCA) to the collected internal state metrics. Since many DBMS metrics are correlated or redundant, directly using the full state vector would increase dimensionality, noise, and optimization difficulty. PCA projects the original state vector into a compact representation while preserving most of its variance, thereby providing a lower-dimensional and more stable state input for refinement.

The output of Phase~0 therefore consists of two workload-specific priors: (1) a Top-$K$ knob set identified through Random-Forest-based reranking, and (2) a compact PCA-based state representation derived from internal metrics. These priors do not directly determine the final configuration. Instead, they constrain and inform the later execution-guided hint-refinement and post-tuning refinement stages, so that PLRTune can reduce unnecessary search while still relying on execution feedback to determine the final tuned values. The concrete sampling budget, knob-selection stability, and state-dimension sensitivity are reported in the experimental section.

\subsection{Phase 1: Execution-Guided Hint Refinement}
\label{sec:phase1}

\begin{figure}[t]
\centering
\includegraphics[width=\columnwidth]{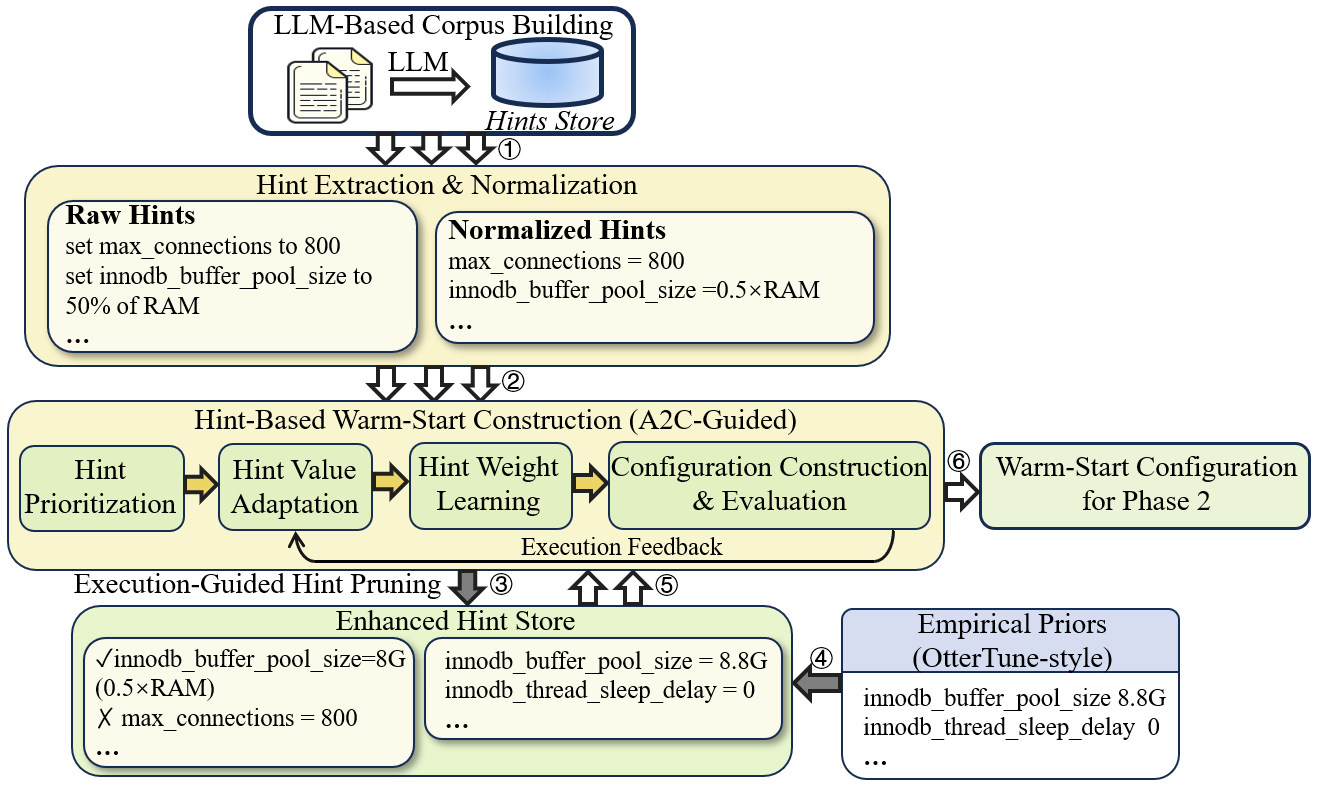}
\captionsetup{skip=2pt}
\caption{Phase~1 workflow for execution-guided hint refinement and initialization construction.}
\label{fig_hint}
\end{figure}

The goal of Phase~1 is to improve initialization quality before post-tuning refinement. Rather than directly applying raw textual hints, PLRTune treats this stage as an execution-guided hint-refinement process. As shown in Figure~\ref{fig_hint}, Phase~1 builds a reusable hint store from external text, converts raw hints into executable formulas, uses execution feedback to adapt hint values and weights, and then refines the hint store through pruning and empirical prior fusion. In this way, Phase~1 turns noisy documentation-derived knowledge into a stronger and more reliable initialization signal for Phase~2.

The upper part of Figure~\ref{fig_hint} corresponds to LLM-based corpus building and hint extraction. PLRTune collects textual tuning knowledge from vendor manuals, engineering blogs, and community discussions, and stores the extracted recommendations in a hint store. Following prior work on text-guided database tuning, the current implementation uses a BERT-based language model to analyze text snippets and identify knob-related recommendations. This step provides a reusable source of prior knowledge, but PLRTune does not directly apply the raw textual hints because they may be incomplete, redundant, or workload-dependent.

The next block in Figure~\ref{fig_hint} performs hint extraction and normalization. Raw hints expressed in natural language are converted into structured parameter--value formulas. A normalized hint may specify either an absolute value, such as setting a connection-related knob to a fixed value, or a resource-relative value, such as setting a memory-related knob as a fraction of RAM. Resource-relative hints are instantiated using system properties such as RAM or CPU resources. This normalization step is important because it turns heterogeneous textual recommendations into executable DBMS configuration rules, allowing the later initializer to operate on structured hints rather than unprocessed text.

The middle part of Figure~\ref{fig_hint} corresponds to hint-based warm-start construction. Starting from the normalized hint set, PLRTune first prioritizes candidate hints and selects a small subset for the current initialization step. It then uses an A2C-guided decision module to adapt the suggested values around their original recommendations and to learn how much weight each selected hint should receive. The adapted and weighted hints are aggregated into an executable DBMS configuration, which is then evaluated under the benchmark workload. The observed objective value is fed back to guide later value adaptation and weight learning. Through this design, Phase~1 improves initialization quality not by blindly trusting textual hints, but by using execution feedback to select, adjust, and combine them within a small number of warm-start steps.

\begin{table*}[t]
\centering
\captionsetup{skip=3pt}
\caption{Representative OtterTune-style recommendations used for hint fusion in Phase~1.}
\label{tab:ottertune-hints}
\begin{tabular}{p{0.42\textwidth} p{0.42\textwidth}}
\hline
\textbf{(a) OtterTune Configuration (MySQL)} & \textbf{(b) OtterTune Configuration (PostgreSQL)} \\
\hline
\texttt{innodb\_buffer\_pool\_size} = 8.8\,G & \texttt{shared\_buffers} = 4\,G \\
\texttt{innodb\_thread\_sleep\_delay} = 0 & \texttt{checkpoint\_segments} = 540 \\
\texttt{innodb\_flush\_method} = \texttt{O\_DIRECT} & \texttt{effective\_cache\_size} = 18\,G \\
\texttt{innodb\_log\_file\_size} = 1.3\,G & \texttt{bgwriter\_lru\_maxpages} = 1000 \\
\texttt{innodb\_thread\_concurrency} = 0 & \texttt{bgwriter\_delay} = 213\,ms \\
\texttt{innodb\_max\_dirty\_pages\_pct\_lwm} = 0 & \texttt{checkpoint\_completion\_target} = 0.8 \\
\texttt{innodb\_read\_ahead\_threshold} = 56 & \texttt{deadlock\_timeout} = 6\,s \\
\texttt{innodb\_adaptive\_max\_sleep\_delay} = 150000 & \texttt{default\_statistics\_target} = 78 \\
\texttt{innodb\_buffer\_pool\_instances} = 8 & \texttt{effective\_io\_concurrency} = 3 \\
\texttt{thread\_cache\_size} = 9 & \texttt{checkpoint\_timeout} = 1\,h \\
\hline
\end{tabular}
\end{table*}

The lower-left feedback path in Figure~\ref{fig_hint} corresponds to execution-guided hint pruning. In practice, not all extracted hints are equally useful: some consistently improve performance, whereas others are rarely selected or are associated with degraded objective values. PLRTune therefore estimates parameter-level marginal contribution from the initialization traces collected in Phase~1. Parameters that are never selected are categorized as \emph{Never Used}. For the remaining parameters, PLRTune compares the average tuning objective between trials in which the parameter is selected and trials in which it is not, and uses the resulting relative lift to categorize each parameter as \emph{Positive}, \emph{Negative}, or \emph{Neutral}. Parameters categorized as \emph{Negative} or \emph{Never Used} are removed from the reusable hint store before later runs. This pruning mechanism reduces the negative impact of noisy or low-value hints, and its effectiveness is evaluated in the component ablation by comparing PLRTune with the variant without hint refinement.

The lower-right part of Figure~\ref{fig_hint} corresponds to empirical prior fusion. In addition to pruning low-value hints, PLRTune augments the refined hint store with structured empirical recommendations. These priors complement documentation-derived hints with empirically effective settings, especially for important knobs. Representative examples for MySQL and PostgreSQL are shown in Table~\ref{tab:ottertune-hints}. The resulting enhanced hint store is constructed separately for the two DBMSs and is reused by later initialization runs.

A practical advantage of Phase~1 is that corpus processing, hint pruning, and empirical prior fusion are decoupled from repeated tuning runs. The full workflow in Figure~\ref{fig_hint} is mainly executed when building or refreshing the enhanced hint store. Once this store has been materialized, subsequent tuning runs can invoke the initializer directly without rebuilding the full refinement pipeline. The output of Phase~1 is therefore twofold: an enhanced reusable hint store and a stronger initial configuration derived from it, which together provide a more reliable starting point for the post-tuning refinement stage in Phase~2.

\subsection{Phase 2: Post-Tuning Refinement}
\label{sec:phase2}

The goal of Phase~2 is to further improve final tuning quality after the strong initialization obtained in Phase~1. Instead of restarting optimization from the DBMS default configuration or exploring the full candidate knob space, PLRTune treats this phase as residual post-tuning refinement. The action space is restricted to the Top-$K$ important knobs selected by the Phase~0 knob-importance analysis, and the state space is represented by the compact PCA-based state features derived from the pre-sampled observations. Therefore, Phase~2 refines the remaining high-value tuning room within a workload-specific reduced search space.

We adopt TD3 as the reinforcement-learning backbone of Phase~2. TD3 is suitable for continuous knob optimization and improves over standard DDPG-style actor--critic methods through twin critics, delayed actor updates, and target-network-based stabilization. In PLRTune, TD3 is used as a state-aware refinement optimizer rather than a from-scratch black-box optimizer. The actor network proposes adjustments only for the Top-$K$ knobs, while knobs outside the reduced subspace inherit their values from the Phase~1 initialization. The critic networks evaluate the resulting state--action pairs using the compact PCA state representation and the tuning objective as the reward signal.

\begin{figure}[t]
  \centering
  \includegraphics[width=\columnwidth]{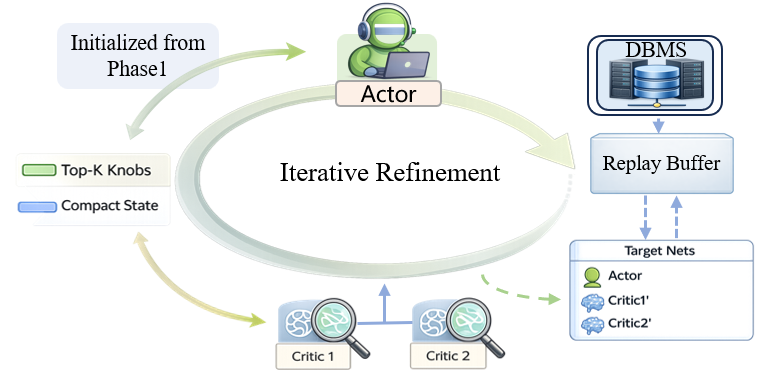}
  \captionsetup{skip=2pt}
  \caption{Structure of Phase~2 post-tuning refinement in PLRTune.}
  \label{fig:td3-structure}
\end{figure}

Figure~\ref{fig:td3-structure} illustrates the iterative refinement process of Phase~2. Starting from the initial configuration produced by Phase~1, the actor generates a candidate update in the reduced Top-$K$ action space. This update is mapped back to an executable DBMS configuration by modifying only the selected important knobs and keeping the remaining knob values unchanged. The configuration is then applied to the DBMS and evaluated under the target workload. During execution, PLRTune collects both the external objective value and the internal state metrics, projects the state metrics into the PCA-based compact state space, and stores the resulting transition in the replay buffer.

The replay buffer supports iterative TD3 updates. In each update, the twin critics estimate the value of candidate state--action pairs, which reduces the risk of overestimating configuration quality. The actor is updated less frequently than the critics, and target networks are used to stabilize training across consecutive tuning steps. These mechanisms make Phase~2 more stable under the limited tuning budget, especially because each DBMS trial is expensive and only a small number of online evaluations can be performed.

This design improves final performance through focused residual exploration. Phase~1 quickly moves the system into a promising region, but textual hints and empirical priors may still leave workload-specific performance unexplored. Phase~2 addresses this gap by using runtime state feedback to continue refinement in the reduced subspace. Since the optimizer no longer needs to search over all candidate knobs or process the full internal metric vector, it can spend the online budget on workload-relevant knobs and compact runtime states.

Together, the three phases provide a compact workload-aware search space, a strong execution-guided initialization, and state-aware post-tuning refinement. This staged design enables PLRTune to achieve both rapid early-stage improvement and high final tuning quality.

\section{Experiments}
\label{sec:experiments}

\subsection{Experimental Setup}
\label{sec:setup}

\subsubsection{Testbed and Implementation Details}
All main experiments are conducted on a single x86-64 KVM virtual machine running Ubuntu~24.04, with 12 vCPUs on an AMD EPYC~7713 host, 64 GB RAM, and a 120 GB SSD. No GPU accelerators are used. The DBMSs are MySQL~8.0.45 and PostgreSQL~16.13, and their validated candidate action spaces and internal metric settings are described in Section~\ref{sec:preliminaries}.

All candidate configurations are evaluated under the same single-machine protocol. For MySQL, selected knob values are applied using \texttt{SET PERSIST}; for PostgreSQL, they are applied using \texttt{ALTER SYSTEM SET}. When a configuration contains restart-required parameters, the DBMS is restarted before benchmark execution so that the configuration takes effect consistently. This protocol standardizes candidate evaluation, while each baseline keeps its original optimization logic as closely as possible.

To evaluate cross-hardware reuse, we use the main testbed as the source server and introduce two target servers: a lower-resource target server with 8 CPU cores, 16 GB memory, and 100 GB storage, and a higher-resource target server with 32 CPU cores, 128 GB memory, and 500 GB storage. All methods are repeated three times for each workload, and we report the average result.

\subsubsection{Benchmarks and Workloads}
We evaluate PLRTune on representative transactional, mixed, and analytical workloads. For MySQL, we use Sysbench read-only, Sysbench write-only, Sysbench read-write, TPC-C, YCSB, and TPC-H. For PostgreSQL, we use Sysbench read-only, Sysbench write-only, and Sysbench read-write. These workloads cover read-intensive, write-intensive, mixed transaction, and analytical access patterns.

Unless otherwise stated, all methods are evaluated within a 100-minute tuning window. For Sysbench, TPC-C, and YCSB, each trial uses a 60-second benchmark execution window. For TPC-H, each trial measures the complete query execution time.

\subsubsection{Baselines}
We compare PLRTune with representative baselines from reinforcement-learning-based, LLM-guided, and sample-efficient tuning methods.

\textbf{HUNTER}~\cite{HUNTER} is a hybrid RL-based tuner that combines GA-generated warm-start samples, Random Forest, PCA, and deep reinforcement learning. Since its original framework benefits from cloned instances and parallelized exploration, we adapt it to the same single-machine evaluation protocol used by all methods.

\textbf{DB-BERT}~\cite{DB-BERT} extracts textual tuning hints from manuals and related documents, and performs hint-guided optimization based on the extracted knowledge.

\textbf{GPTuner}~\cite{GPTuner} is an LLM-assisted tuner that leverages heterogeneous textual knowledge, workload-aware knob selection, and Coarse-to-Fine Bayesian optimization.

\textbf{LlamaTune}~\cite{kanellis2022llamatune} improves sample efficiency through search-space shaping and dimensionality reduction. In our experiments, we instantiate it with SMAC~\cite{hutter2011sequential}.

Each baseline follows the default or recommended algorithm-specific hyperparameters from its original paper or official implementation. We only adapt the execution protocol to ensure a unified single-machine comparison, including the same workload execution window, configuration application procedure, and 100-minute tuning budget. For PLRTune, the Phase~1 initialization settings are aligned with DB-BERT where applicable, and the Phase~2 TD3-related settings are aligned with HUNTER where applicable.

\subsubsection{Evaluation Metrics}
For transactional and mixed workloads, we evaluate both throughput and tail latency using the scalarized objective
\[ \mathrm{fitness}=\frac{\mathrm{TPS}}{\mathrm{p95}}, \]
where a higher value indicates better performance. This objective favors configurations that improve throughput without excessively increasing tail latency. For TPC-H, we use execution time as the objective, where a lower value is better.

We also analyze convergence behavior with respect to wall-clock tuning time and compare the time required to reach strong or best-observed configurations under the same evaluation protocol.

\subsection{Phase~0 Prior Construction Analysis}

This section analyzes two key parameter choices in Phase~0: the pre-sampling budget used for stable important-knob identification, and the retained PCA state dimension used for compact state representation. These analyses determine the Phase~0 settings used in the subsequent experiments.

\subsubsection{Knob Stability Under Increasing Pre-Sampling Steps}

We first study how many pre-sampling steps are needed to obtain a stable Top-$K$ important-knob set. For a given prefix length (n), we compute the Top-$K$ knob set from the first (n) pre-sampling steps and compare it with the Top-$K$ set obtained from the full pre-sampling collection. The similarity is measured by the Jaccard index between $\mathrm{TopK}(n)$ and $\mathrm{TopK}(\mathrm{full})$. A higher Jaccard value indicates that the important-knob set obtained from the prefix samples is closer to the final Top-$K$ result.

\begin{figure}[t]
\centering
\includegraphics[width=\columnwidth]{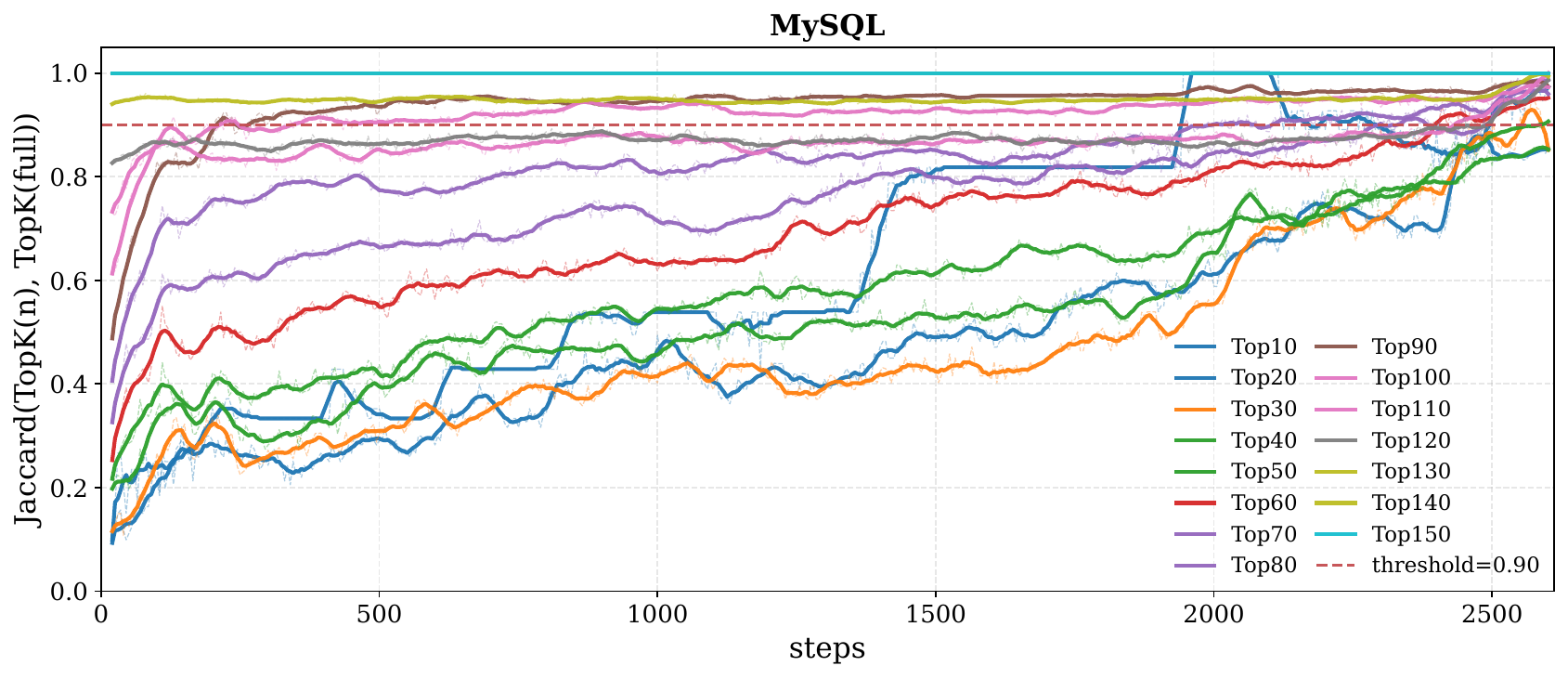}

\vspace{0.25em}

\includegraphics[width=\columnwidth]{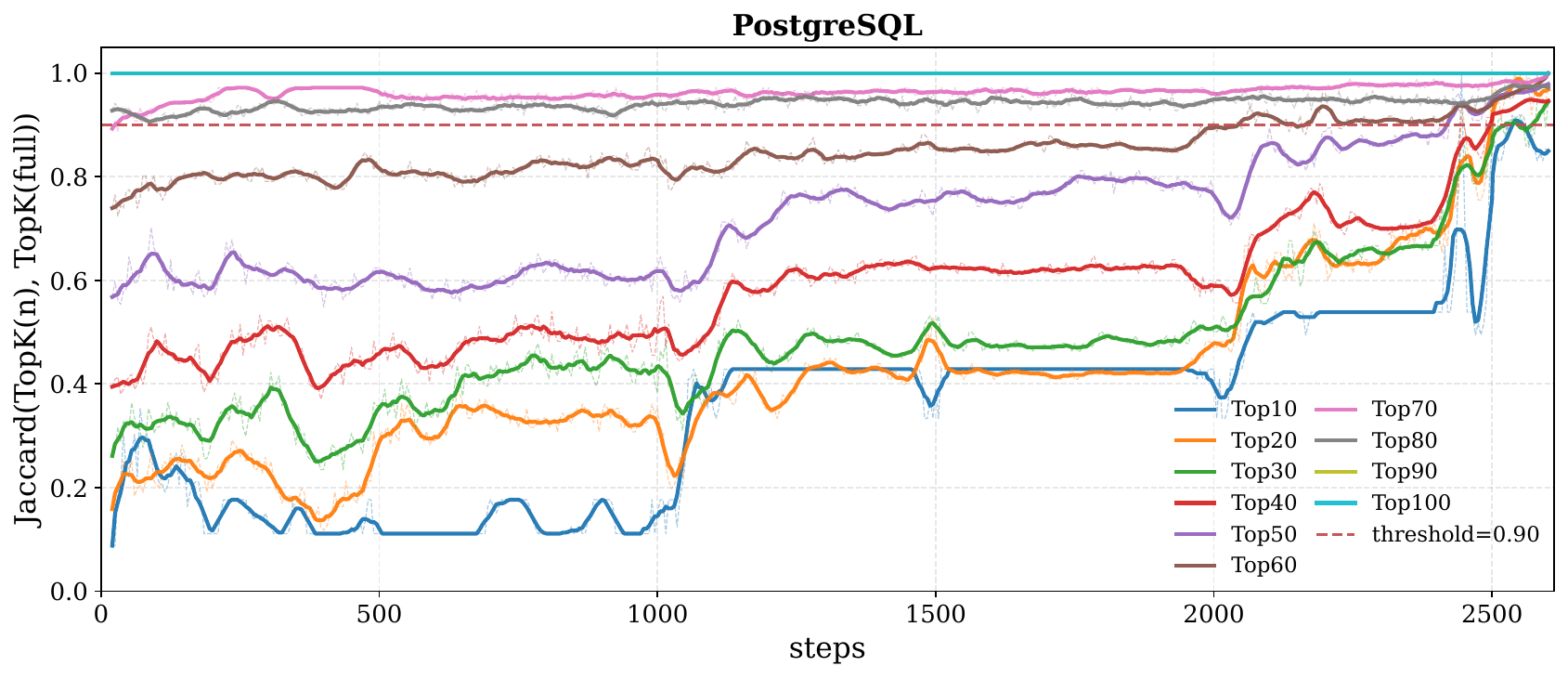}
\captionsetup{skip=2pt}
\caption{Top-$K$ stability under increasing pre-sampling steps on MySQL and PostgreSQL.}
\label{fig:topk-stability}
\end{figure}

Figure~\ref{fig:topk-stability} shows the stability curves for both MySQL and PostgreSQL under different Top-$K$ settings. In both systems, the Jaccard similarity increases steadily as the number of pre-sampling steps grows, indicating that the identified important-knob sets gradually converge to the full-sample result. By around 2500 pre-sampling steps, most curves approach or exceed the 0.90 threshold, suggesting that the Top-$K$ results have become sufficiently stable.

Based on this observation, we set the Phase~0 pre-sampling budget to 2500 steps in PLRTune. This setting provides stable workload-specific knob priors for later stages without introducing unnecessary additional pre-sampling cost.

\subsubsection{PCA-Based State Compression}

\begin{figure}[t]
\centering
\begin{minipage}{0.48\columnwidth}
\centering
\includegraphics[width=\linewidth]{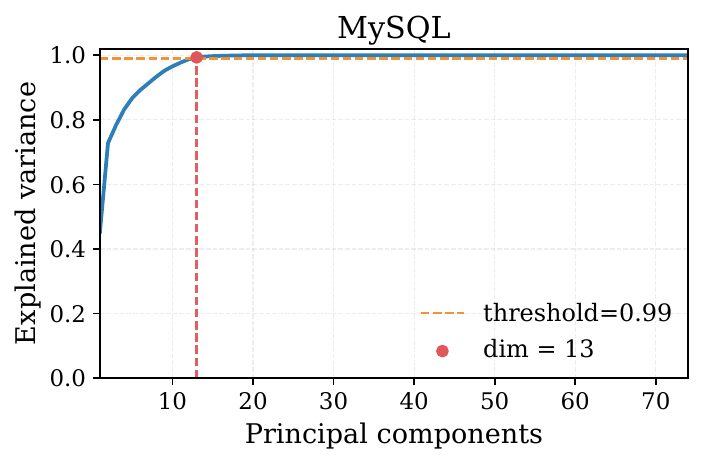}
\end{minipage}
\hfill
\begin{minipage}{0.48\columnwidth}
\centering
\includegraphics[width=\linewidth]{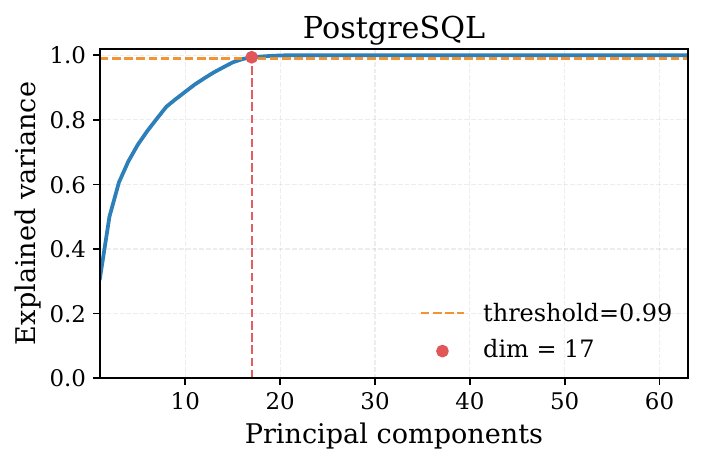}
\end{minipage}
\vspace{-0.2em}
\captionsetup{skip=3pt}
\caption{Cumulative explained variance under PCA for MySQL and PostgreSQL.}
\label{fig:pca-state-compression}
\end{figure}

We next determine the retained state dimension after PCA-based compression. Figure~\ref{fig:pca-state-compression} reports the cumulative explained variance as the number of principal components increases for MySQL and PostgreSQL. We select the retained dimension using a cumulative explained variance threshold of 0.99, so that the compressed state representation preserves approximately 99\% of the variance in the original internal metrics.

Under this criterion, the retained state dimension is 13 for MySQL and 17 for PostgreSQL. These dimensions are used as the compact state representation in the subsequent Phase~2 refinement experiments.

\subsection{Phase~1 Convergence Behavior}
\label{sec:phase1-convergence}

This section determines the tuning-step budget used for Phase~1 in the final PLRTune configuration. To isolate the behavior of the hint-guided initialization stage, we run Phase~1 alone on three representative Sysbench workloads for both MySQL and PostgreSQL, without enabling the subsequent TD3-based refinement.

\begin{figure}[t]
  \centering
  \captionsetup[subfigure]{labelformat=empty}
    \setlength{\tabcolsep}{1.5pt}
  \begin{tabular}{ccc}
    \includegraphics[width=0.32\columnwidth,height=0.10\textheight,keepaspectratio,trim=6 6 6 6,clip]{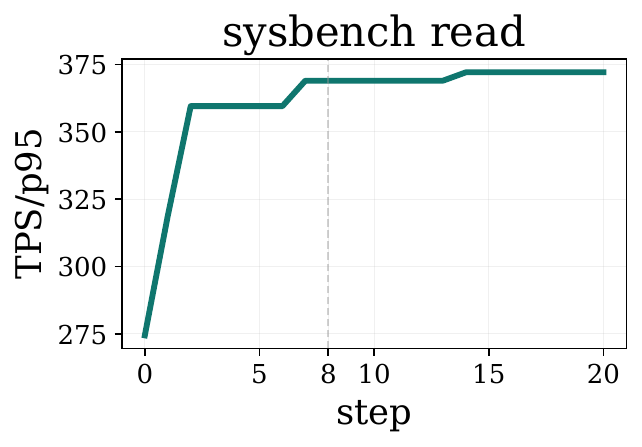} &
    \includegraphics[width=0.32\columnwidth,height=0.10\textheight,keepaspectratio,trim=6 6 6 6,clip]{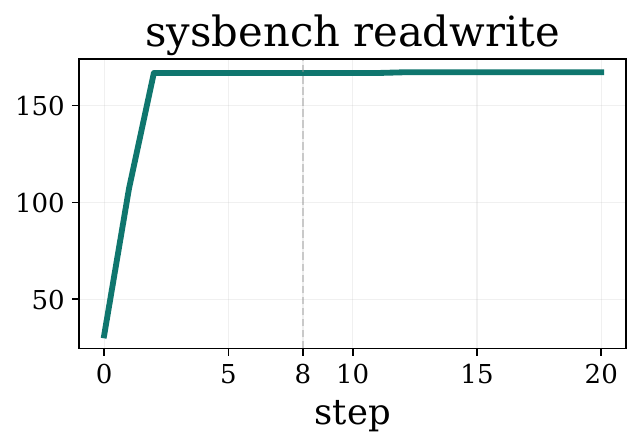} &
    \includegraphics[width=0.32\columnwidth,height=0.10\textheight,keepaspectratio,trim=6 6 6 6,clip]{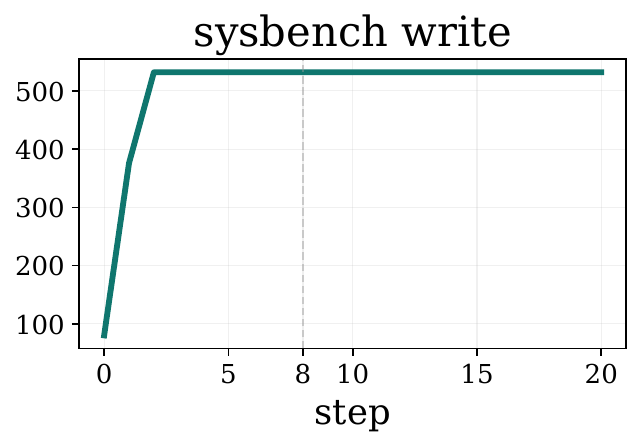} \\[-0.2mm]

    \multicolumn{3}{c}{\small (a) MySQL workloads} \\[0.6mm]

    \includegraphics[width=0.32\columnwidth,height=0.10\textheight,keepaspectratio,trim=6 6 6 6,clip]{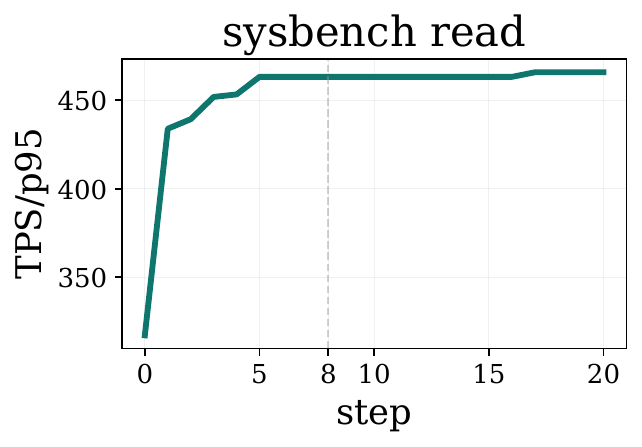} &
    \includegraphics[width=0.32\columnwidth,height=0.10\textheight,keepaspectratio,trim=6 6 6 6,clip]{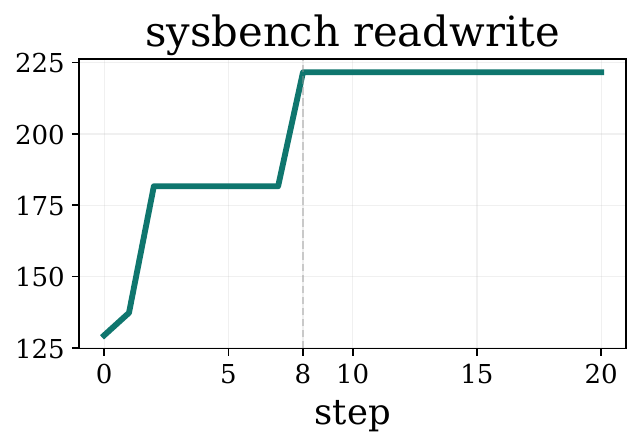} &
    \includegraphics[width=0.32\columnwidth,height=0.10\textheight,keepaspectratio,trim=6 6 6 6,clip]{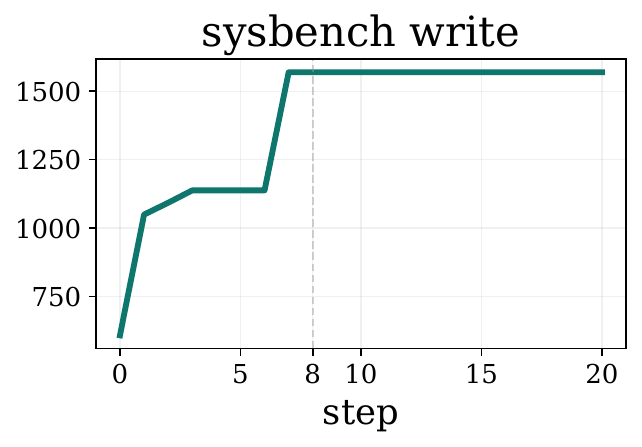} \\[-0.2mm]

    \multicolumn{3}{c}{\small (b) PostgreSQL workloads}
  \end{tabular}

  \captionsetup{skip=2pt}
  \caption{Standalone convergence behavior of the Phase~1 hint-guided initialization stage on three representative Sysbench workloads for MySQL and PostgreSQL.}
  \label{fig:phase1-convergence}
\end{figure}

Figure~\ref{fig:phase1-convergence} shows that Phase~1 improves rapidly during the first few tuning steps across all six workloads. The curves become much flatter after several additional iterations, indicating that most initialization gains have already been obtained. Although a few workloads still show small improvements after step~8, the marginal gain becomes limited compared with the additional tuning cost.

Based on this observation, we set the Phase~1 budget to 8 tuning steps in the final PLRTune configuration. This setting provides a strong initial configuration for Phase~2 while avoiding spending too much of the end-to-end tuning budget on the initialization stage. The remaining budget is then used by Phase~2 for TD3-based post-tuning refinement.

\subsection{End-to-End Tuning Performance}
\label{sec:end2end}

\begin{figure*}[t]
  \centering
  \captionsetup[subfigure]{labelformat=empty}
  \graphicspath{{C:/Users/96449/Desktop/L2TTune/results/pdf/runtime_100min/}}

  \includegraphics[width=0.7\textwidth,trim=4 4 4 4,clip]{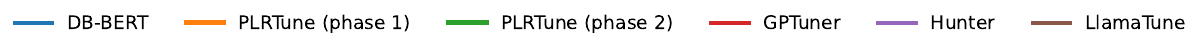}

  \vspace{0.4mm}

  \begin{tabular}{ccc}
    \includegraphics[width=0.31\textwidth,height=0.10\textheight,keepaspectratio,trim=6 6 6 6,clip]{\detokenize{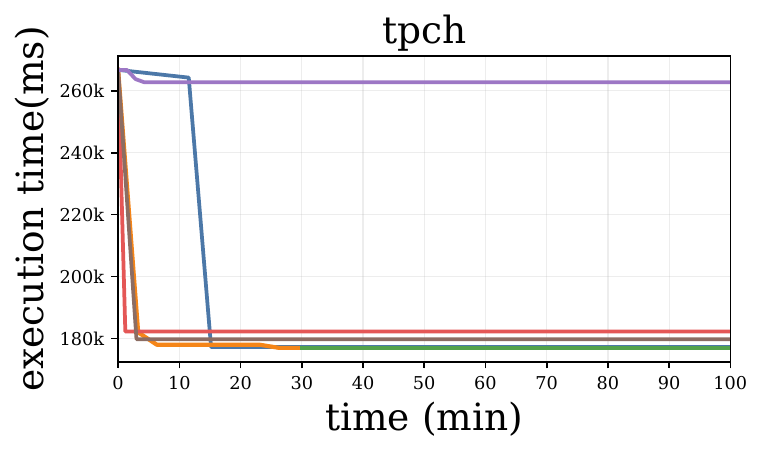}} &
    \includegraphics[width=0.31\textwidth,height=0.10\textheight,keepaspectratio,trim=6 6 6 6,clip]{\detokenize{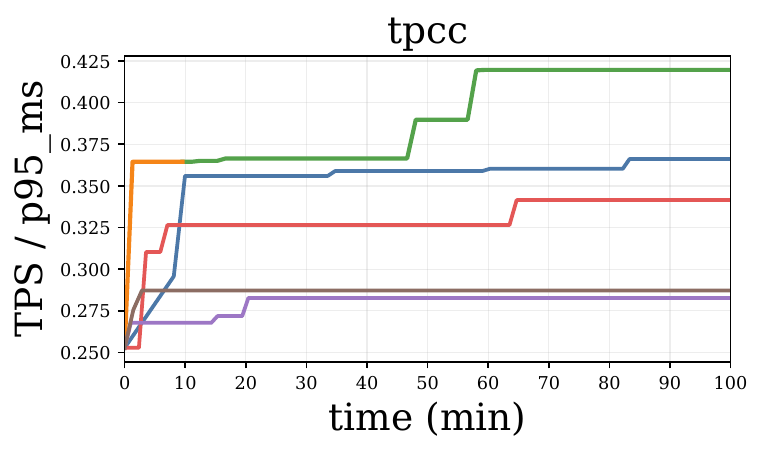}} &
    \includegraphics[width=0.31\textwidth,height=0.10\textheight,keepaspectratio,trim=6 6 6 6,clip]{\detokenize{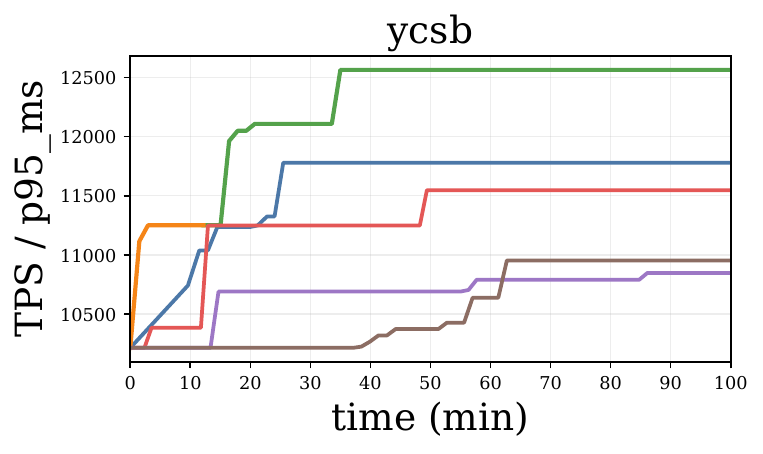}} \\[0.4mm]

    \includegraphics[width=0.31\textwidth,height=0.10\textheight,keepaspectratio,trim=6 6 6 6,clip]{\detokenize{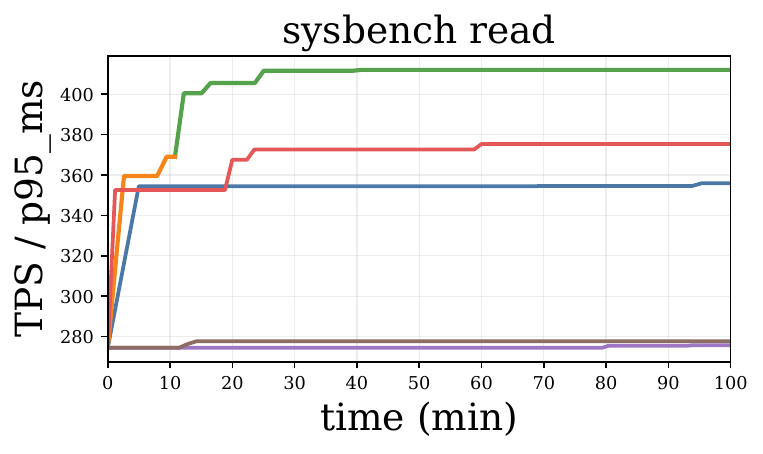}} &
    \includegraphics[width=0.31\textwidth,height=0.10\textheight,keepaspectratio,trim=6 6 6 6,clip]{\detokenize{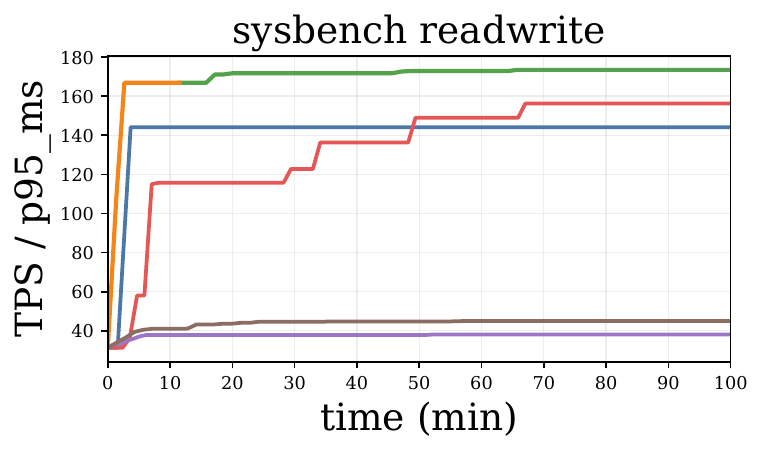}} &
    \includegraphics[width=0.31\textwidth,height=0.10\textheight,keepaspectratio,trim=6 6 6 6,clip]{\detokenize{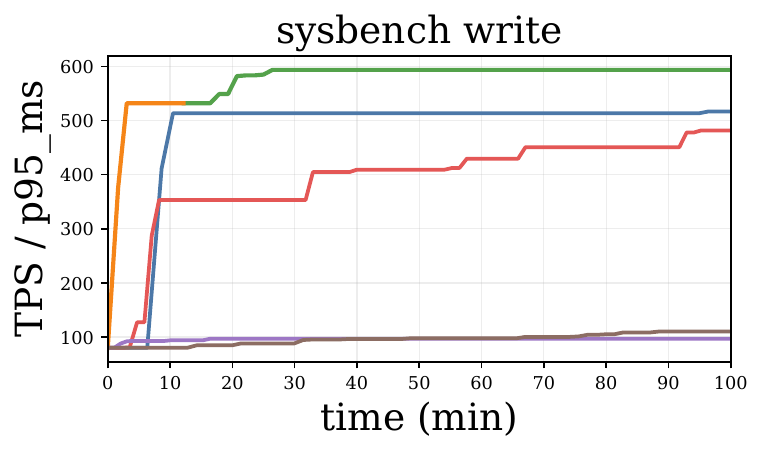}} \\[-0.2mm]

    \multicolumn{3}{c}{\small (a) MySQL workloads} \\[0.5mm]

    \includegraphics[width=0.31\textwidth,height=0.10\textheight,keepaspectratio,trim=6 6 6 6,clip]{\detokenize{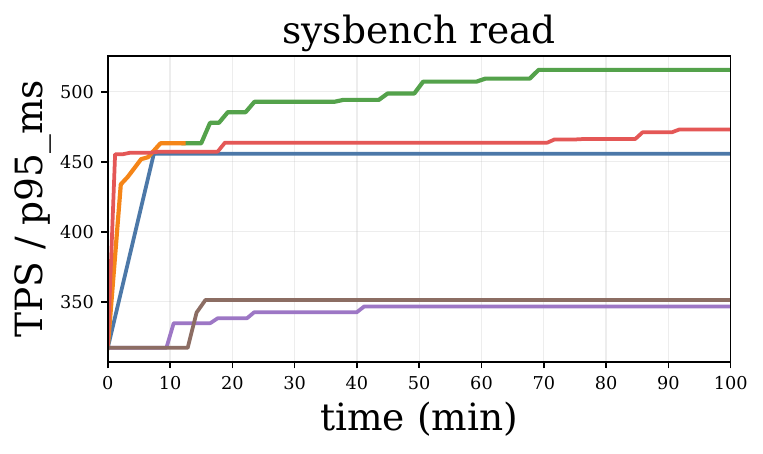}} &
    \includegraphics[width=0.31\textwidth,height=0.10\textheight,keepaspectratio,trim=6 6 6 6,clip]{\detokenize{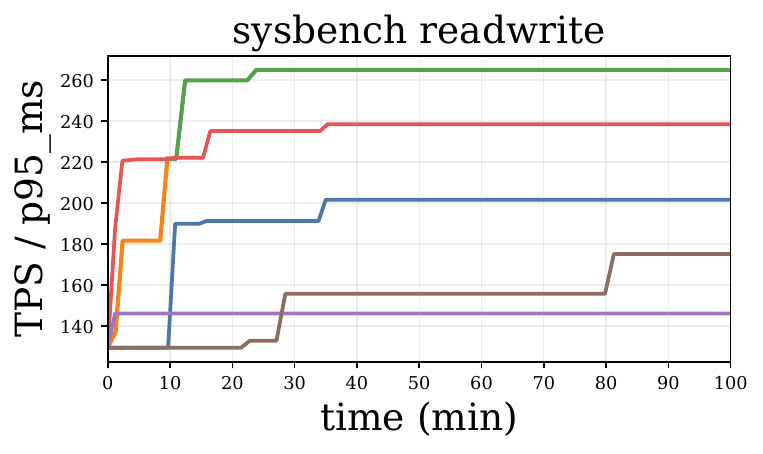}} &
    \includegraphics[width=0.31\textwidth,height=0.10\textheight,keepaspectratio,trim=6 6 6 6,clip]{\detokenize{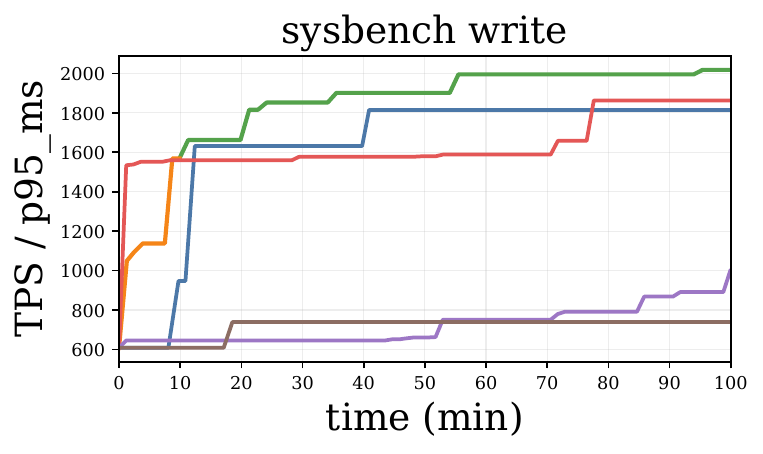}} \\[-0.2mm]

    \multicolumn{3}{c}{\small (b) PostgreSQL workloads}
  \end{tabular}

  \captionsetup{skip=2pt}
  \caption{End-to-end best-so-far tuning performance of PLRTune and representative baselines across MySQL and PostgreSQL workloads.}
  \label{fig:all-main-performance}
\end{figure*}

\subsubsection{Overall Results}

In this section, we evaluate the end-to-end tuning performance of PLRTune on both MySQL and PostgreSQL workloads under the 100-minute tuning budget. All results use the final PLRTune configuration: Phase~1 performs 8 hint-guided initialization steps, and Phase~2 conducts TD3-based post-tuning refinement over the Top-50 workload-relevant knobs identified in Phase~0. The choice of Top-50 is analyzed later in Section~\ref{sec:topn-sensitivity}.

Figure~\ref{fig:all-main-performance} shows the best-so-far tuning curves with respect to wall-clock time, including all compared methods. Table~\ref{tab:end-to-end-summary} summarizes the best-baseline comparison for each workload by reporting the best-performing alternative, its best objective value and time-to-best, the time required by PLRTune to reach that alternative's best performance level, PLRTune's own best objective value and time-to-best, and the final gain over the best-performing alternative. Overall, PLRTune shows a clear two-stage trajectory: Phase~1 quickly moves the DBMS to a strong initial region, while Phase~2 continues to improve the best-so-far performance within the reduced Top-50 knob subspace. This confirms that hint-guided initialization provides a strong start, but further state-aware refinement is still necessary to exploit the remaining optimization room.

\begin{table*}[t]
\centering
\caption{End-to-end best-so-far and runtime-efficiency comparison against the best-performing alternative within the 100-minute budget.}
\label{tab:end-to-end-summary}
\scriptsize
\setlength{\tabcolsep}{3pt}
\resizebox{\textwidth}{!}{%
\begin{tabular}{llccccccc}
\toprule
\textbf{Workload} &
\textbf{Metric} &
\shortstack{\textbf{Best-Performing}\\\textbf{Alternative}} &
\shortstack{\textbf{Alternative}\\\textbf{Best}} &
\shortstack{\textbf{Alternative}\\\textbf{Time (min)}} &
\shortstack{\textbf{PLRTune Target Time}\\\textbf{(min; Runtime Change)}} &
\shortstack{\textbf{PLRTune}\\\textbf{Best}} &
\shortstack{\textbf{PLRTune Best Time}\\\textbf{(min; Runtime Change)}} &
\shortstack{\textbf{Final}\\\textbf{Gain}} \\
\midrule
MySQL TPC-H & Exec. time (ms) & DB-BERT & 177239.00 & 61.15 & 26.29 (2.33$\times$ faster) & \textbf{176960.00} & 26.29 (2.33$\times$ faster) & \textbf{+0.16\%} \\
MySQL TPC-C & TPS/p95 & DB-BERT & 0.3662 & 83.33 & 16.68 (5.00$\times$ faster) & \textbf{0.4197} & 59.45 (1.40$\times$ faster) & \textbf{+14.61\%} \\
MySQL YCSB & TPS/p95 & DB-BERT & 11778.88 & 25.48 & 16.46 (1.55$\times$ faster) & \textbf{12563.14} & 34.99 (1.37$\times$ longer) & \textbf{+6.66\%} \\
MySQL Read & TPS/p95 & GPTuner & 375.34 & 59.98 & 12.21 (4.91$\times$ faster) & \textbf{412.07} & 40.72 (1.47$\times$ faster) & \textbf{+9.79\%} \\
MySQL Read-Write & TPS/p95 & GPTuner & 156.21 & 67.04 & 2.67 (25.11$\times$ faster) & \textbf{173.40} & 65.68 (1.02$\times$ faster) & \textbf{+11.01\%} \\
MySQL Write & TPS/p95 & DB-BERT & 516.71 & 96.44 & 3.04 (31.72$\times$ faster) & \textbf{593.48} & 26.43 (3.65$\times$ faster) & \textbf{+14.86\%} \\
PG Read & TPS/p95 & GPTuner & 473.10 & 91.74 & 16.42 (5.59$\times$ faster) & \textbf{515.69} & 69.17 (1.33$\times$ faster) & \textbf{+9.00\%} \\
PG Read-Write & TPS/p95 & GPTuner & 238.42 & 35.28 & 12.42 (2.84$\times$ faster) & \textbf{264.90} & 23.83 (1.48$\times$ faster) & \textbf{+11.11\%} \\
PG Write & TPS/p95 & GPTuner & 1863.31 & 77.63 & 35.54 (2.18$\times$ faster) & \textbf{2018.12} & 95.41 (1.23$\times$ longer) & \textbf{+8.31\%} \\
\midrule
\textbf{Average} & -- & -- & -- & -- & \textbf{9.03$\times$ faster} & -- & -- & \textbf{+9.50\%} \\
\bottomrule
\end{tabular}%
}
\end{table*}

As reported in Table~\ref{tab:end-to-end-summary}, PLRTune achieves the best final result on all nine workloads within the 100-minute tuning budget. On MySQL, PLRTune improves over the best-performing alternative by \textbf{14.61\%} on TPC-C, \textbf{6.66\%} on YCSB, \textbf{9.79\%} on Sysbench read, \textbf{11.01\%} on Sysbench read-write, and \textbf{14.86\%} on Sysbench write, with an average improvement of \textbf{11.39\%} over the five TPS/p95 workloads. On PostgreSQL, PLRTune improves over the best-performing alternative by \textbf{9.00\%} on Sysbench read, \textbf{11.11\%} on Sysbench read-write, and \textbf{8.31\%} on Sysbench write, with an average improvement of \textbf{9.47\%}. Across all workloads, PLRTune improves over the corresponding best-performing alternatives by \textbf{9.50\%} on average, and by \textbf{10.67\%} on average over the eight TPS/p95 workloads.

The curves in Figure~\ref{fig:all-main-performance} show a consistent pattern across both DBMSs. PLRTune, DB-BERT, and GPTuner often obtain strong early gains, confirming the usefulness of hint-guided or knowledge-guided initialization. However, DB-BERT and GPTuner frequently become flat after their initial improvements, whereas PLRTune continues to improve after Phase~1. This indicates that textual guidance can locate a promising region quickly, but cannot fully exploit the remaining tuning room. PLRTune further converts this residual room into final performance gains through TD3-based refinement in the reduced knob subspace. In contrast, HUNTER and LlamaTune generally improve more slowly under our single-machine evaluation protocol, especially because they do not provide the same hint-guided initialization mechanism as PLRTune.

TPC-H is the only workload where the final gap is small, with PLRTune improving over the best-performing alternative by only \textbf{0.16\%}. Most methods except HUNTER converge rapidly on this workload and reach a strong region early, leaving limited residual tuning room after the initial stage.

Taken together, these results show that PLRTune consistently improves both tuning quality and convergence behavior across different DBMSs and workload types. Its main advantage comes from combining fast hint-guided initialization with reduced-space TD3 refinement: Phase~1 quickly reaches a strong region, while Phase~2 continues to exploit the remaining optimization room that documentation-guided baselines do not fully close.

\subsubsection{Runtime Efficiency Analysis}
\label{sec:runtime-efficiency}

We further analyze runtime efficiency from an end-to-end tuning perspective using the same best-so-far curves as in the main experiments. The measured wall-clock time includes configuration application, workload execution, and metric collection. As shown in Table~\ref{tab:end-to-end-summary}, PLRTune reaches the best-performing alternative's best performance level earlier on every workload. Averaged across workloads, PLRTune reaches this shared target \textbf{9.03$\times$ faster}, showing that it can obtain baseline-level tuning quality much earlier under the same end-to-end evaluation protocol.

We also report PLRTune's own-best time as a stricter criterion, because PLRTune's own best performance is higher than the best-performing alternative's best performance on every workload. In most workloads, PLRTune reaches its own stronger best result faster than the best-performing alternative reaches its own best. The only exceptions are MySQL YCSB and PostgreSQL Sysbench write, where PLRTune spends more time reaching its own best result. However, in both cases, PLRTune has already reached the alternative's best-performance target earlier, and the additional time is spent improving beyond that target. Overall, PLRTune reaches baseline-level performance faster on all workloads while achieving an average final improvement of \textbf{9.50\%}, with the maximum improvement reaching \textbf{14.86\%}.

\subsection{Ablation and Sensitivity Analysis}

\subsubsection{Top-N Sensitivity}
\label{sec:topn-sensitivity}

\begin{figure*}[t]
  \centering
  \captionsetup[subfigure]{labelformat=empty}
  \setlength{\tabcolsep}{1.5pt}

  \begin{tabular}{ccc}
    \includegraphics[width=0.32\textwidth,height=0.12\textheight,keepaspectratio,trim=6 6 6 6,clip]{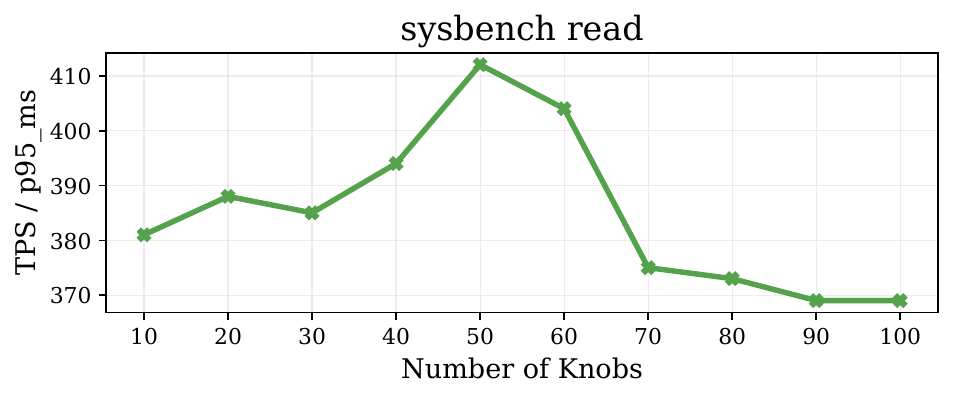} &
    \includegraphics[width=0.32\textwidth,height=0.12\textheight,keepaspectratio,trim=6 6 6 6,clip]{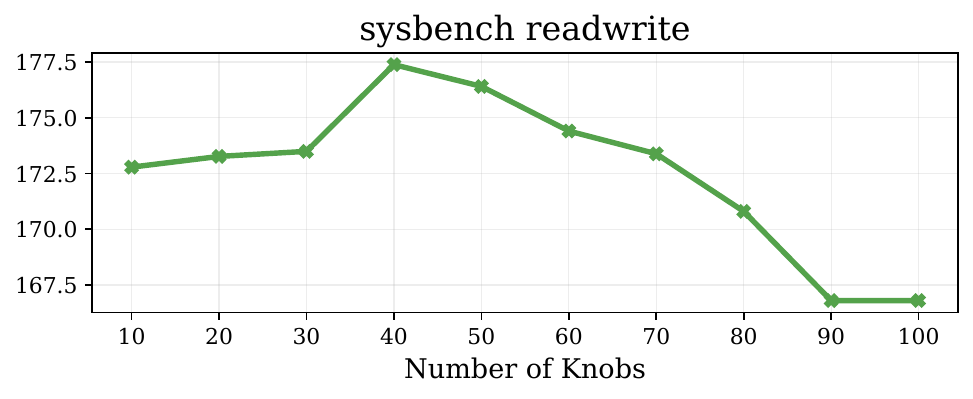} &
    \includegraphics[width=0.32\textwidth,height=0.12\textheight,keepaspectratio,trim=6 6 6 6,clip]{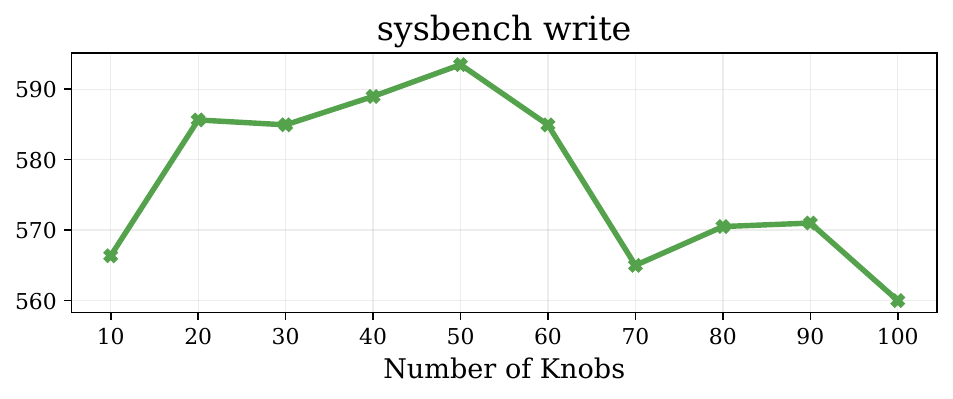} \\[-0.2mm]

    \multicolumn{3}{c}{\small (a) MySQL workloads} \\[0.6mm]

    \includegraphics[width=0.32\textwidth,height=0.12\textheight,keepaspectratio,trim=6 6 6 6,clip]{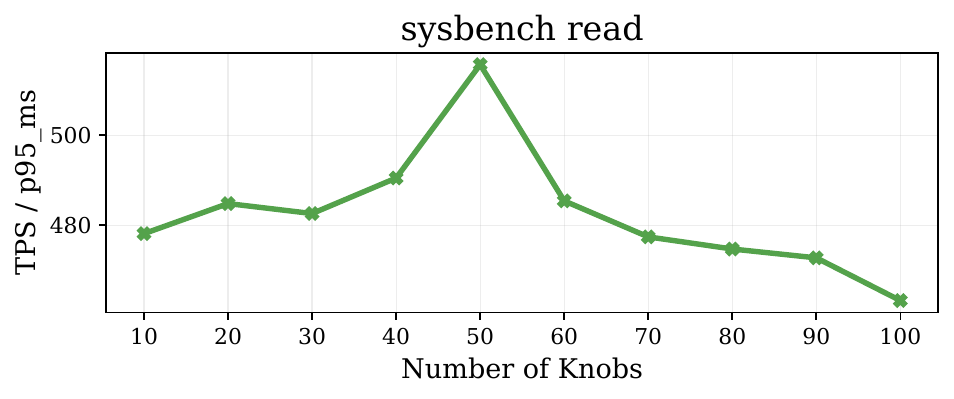} &
    \includegraphics[width=0.32\textwidth,height=0.12\textheight,keepaspectratio,trim=6 6 6 6,clip]{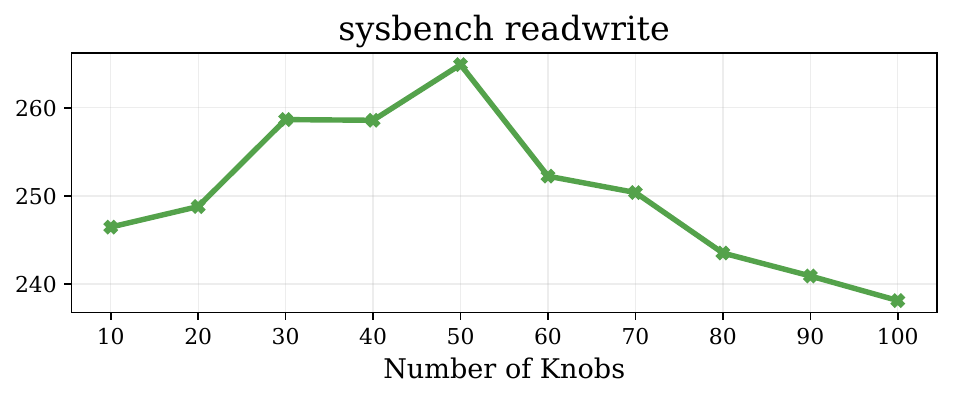} &
    \includegraphics[width=0.32\textwidth,height=0.12\textheight,keepaspectratio,trim=6 6 6 6,clip]{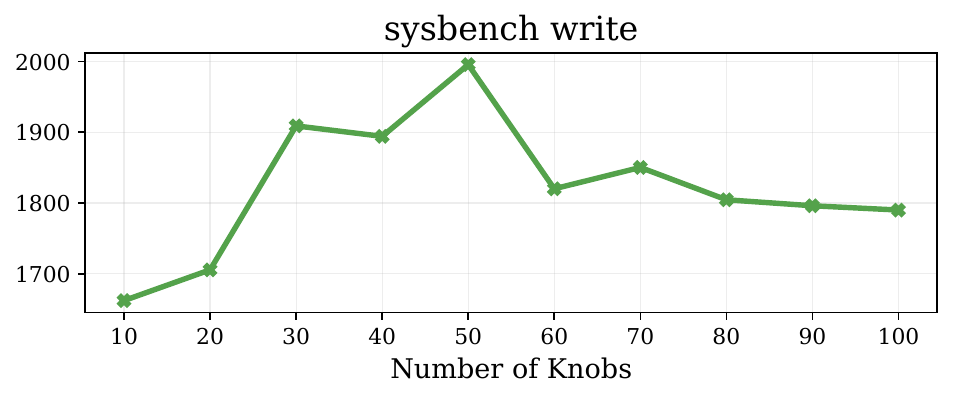} \\[-0.2mm]

    \multicolumn{3}{c}{\small (b) PostgreSQL workloads}
  \end{tabular}

  \captionsetup{skip=2pt}
  \caption{Top-$N$ sensitivity under the 100-minute tuning window on three Sysbench workloads for MySQL and PostgreSQL.}
  \label{fig:topn-sensitivity-all}
\end{figure*}

We next study the sensitivity of PLRTune to the size of the reduced knob subspace. For both MySQL and PostgreSQL, we evaluate Top-$N$ settings from 10 to 100 on three Sysbench workloads each under the same 100-minute tuning window used in the main end-to-end experiments. Each variant follows the same PLRTune pipeline and changes only the number of important knobs retained for Phase~2 refinement.

As shown in Figure~\ref{fig:topn-sensitivity-all}, Top-50 reaches the best or near-best result on almost all six workloads. Smaller settings such as Top-10 or Top-20 often exclude useful knobs and limit the final achievable performance, whereas larger settings reintroduce more low-value or noisy dimensions and bring little additional benefit. These results show that Top-50 provides a robust balance between search efficiency and refinement capacity across both DBMSs. Based on these observations, we adopt Top-50 as the default reduced knob subspace in the final PLRTune configuration.

\subsubsection{Phase and Component Ablation}

\begin{figure}[t]
  \centering
  \includegraphics[width=0.8\columnwidth]{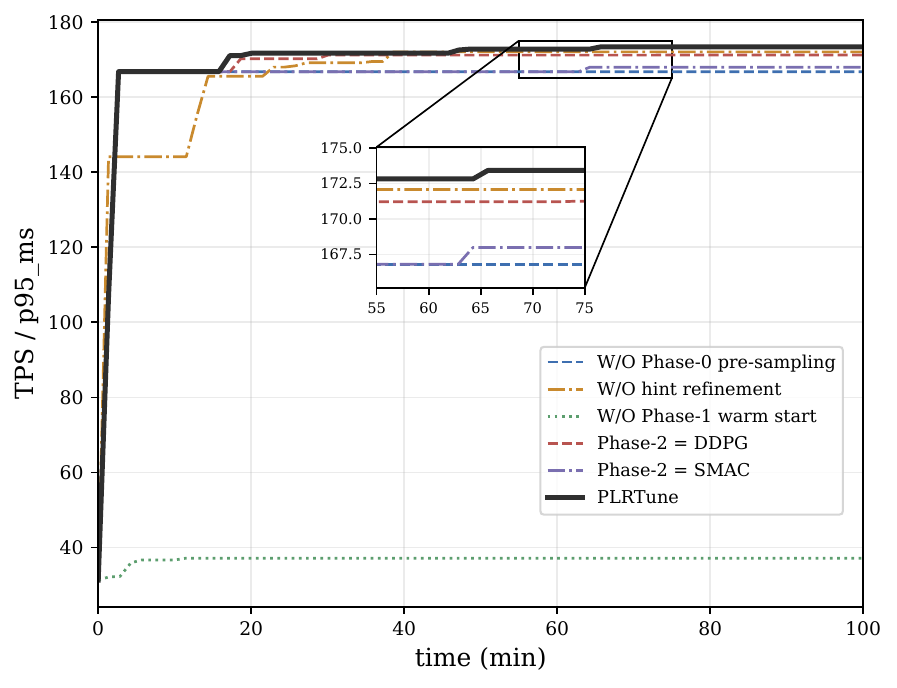}
  \captionsetup{skip=2pt}
  \caption{Ablation results on MySQL Sysbench read-write.}
  \label{fig:ablation-rw}
\end{figure}

We next analyze the contribution of each major component of PLRTune using MySQL Sysbench read-write as a representative controlled ablation workload. The evaluated variants are designed to examine the three core roles of PLRTune: Phase~0 reduces the online search burden through workload-specific reduced-subspace construction, Phase~1 improves initialization quality through execution-guided hint refinement, and Phase~2 improves final tuning quality through state-aware post-tuning refinement. Figure~\ref{fig:ablation-rw} compares the full PLRTune with variants that remove Phase~0 reduced-subspace construction, remove hint refinement, remove Phase~1 initialization, or replace the Phase~2 optimizer with DDPG or SMAC. The full PLRTune achieves the strongest best-so-far performance, while all ablated variants show clear degradation.

Removing Phase~0 weakens later refinement because the optimizer no longer benefits from a workload-specific reduced knob prior or compact PCA-based state representation. Removing hint refinement leads to a weaker Phase~1 initialization, and removing Phase~1 causes much slower early-stage progress, making the optimizer closer to cold-start RL. These results confirm that both reduced-subspace construction and execution-guided warm start are important for effective refinement.

Replacing TD3 with DDPG or SMAC also lowers the final result. This shows that Phase~2 is not merely a second search stage, but benefits from state-aware refinement in the reduced subspace. In particular, the weaker SMAC variant does not imply that SMAC is ineffective as a general-purpose optimizer; rather, it reflects the specific role of Phase~2 in PLRTune, where the goal is residual refinement from a strong initialization using compact runtime state feedback. Overall, the strongest result is achieved only when workload-specific reduced-subspace construction, enhanced hint-guided initialization, and TD3-based post-tuning refinement are used together.

\subsection{Cross-Hardware Reuse Evaluation}
\label{sec:transfer}

We evaluate whether the workload-specific important-knob priors identified by PLRTune remain useful when the same workload type is deployed on differently configured servers. We consider two heterogeneous target servers: a lower-resource target server with 8 CPU cores, 16 GB memory, and 100 GB storage, and a higher-resource target server with 32 CPU cores, 128 GB memory, and 500 GB storage.

\begin{figure}[t]
  \centering

  \includegraphics[width=0.92\columnwidth]{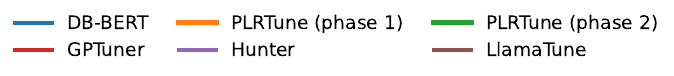}\vspace{0.2em}

  \begin{minipage}{0.32\columnwidth}
    \centering
    \includegraphics[width=\linewidth]{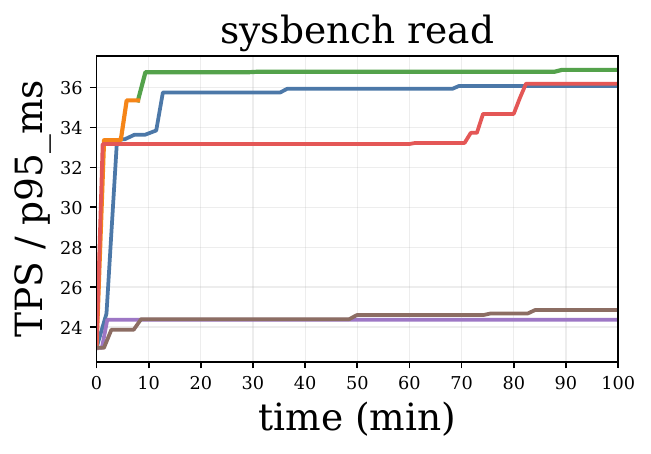}
  \end{minipage}
  \hfill
  \begin{minipage}{0.32\columnwidth}
    \centering
    \includegraphics[width=\linewidth]{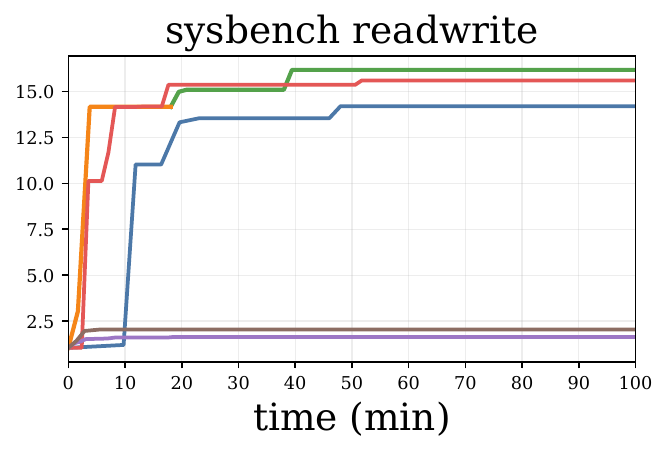}
  \end{minipage}
  \hfill
  \begin{minipage}{0.32\columnwidth}
    \centering
    \includegraphics[width=\linewidth]{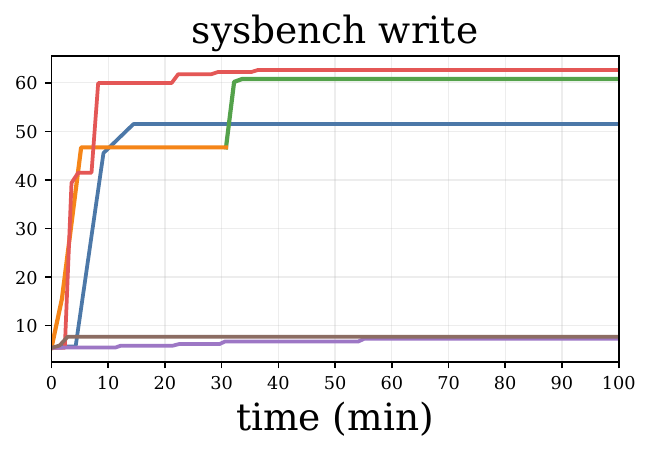}
  \end{minipage}

  \vspace{0.15em}

  {\footnotesize (a) Lower-resource target server}

  \vspace{0.25em}

  \begin{minipage}{0.32\columnwidth}
    \centering
    \includegraphics[width=\linewidth]{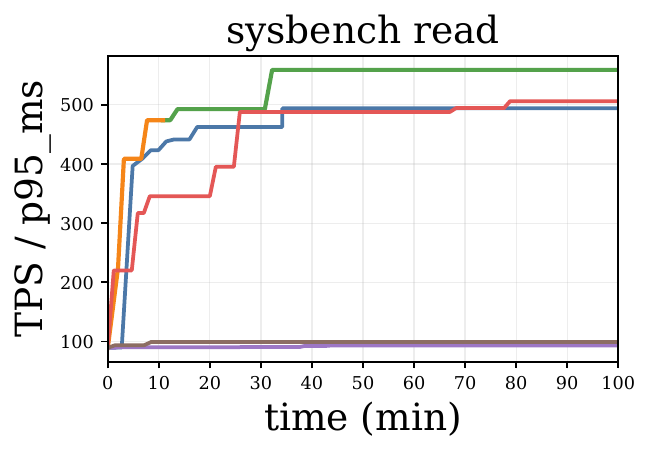}
  \end{minipage}
  \hfill
  \begin{minipage}{0.32\columnwidth}
    \centering
    \includegraphics[width=\linewidth]{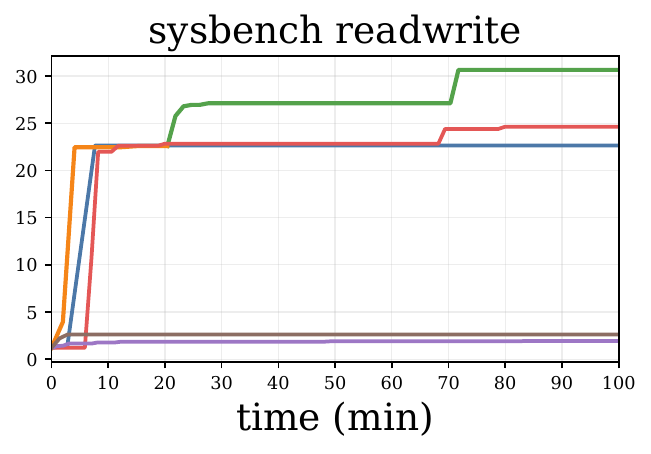}
  \end{minipage}
  \hfill
  \begin{minipage}{0.32\columnwidth}
    \centering
    \includegraphics[width=\linewidth]{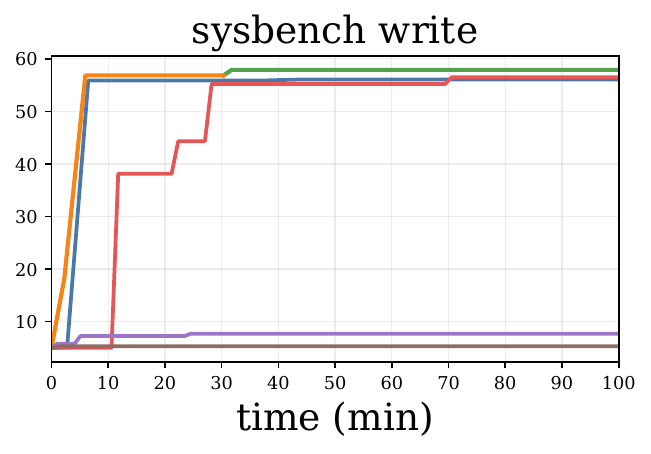}
  \end{minipage}

  \vspace{0.15em}

  {\footnotesize (b) Higher-resource target server}

  \vspace{-0.35em}
  \captionsetup{skip=3pt}
  \caption{Cross-hardware reuse results on lower-resource and higher-resource target servers across three Sysbench workloads.}
  \label{fig:cross-hardware-transfer}
\end{figure}

As shown in Figure~\ref{fig:cross-hardware-transfer}, PLRTune remains effective after the hardware configuration changes. On both the lower-resource and higher-resource target servers, it achieves the best final tuning performance across the tested Sysbench workloads. The two-stage pattern also remains visible: Phase~1 rapidly moves the system to a strong initial region, and Phase~2 further improves the final result through post-tuning refinement.

Importantly, PLRTune reuses only the workload-specific Top-$K$ knob prior identified during Phase~0, rather than directly transferring a final configuration from the source server. On each target server, this prior defines the reduced tuning subspace, while Phase~1 and Phase~2 are still executed locally to adapt the configuration to the new hardware environment. These results indicate that the workload-specific priors identified by PLRTune remain useful beyond a single fixed hardware setting.

\section{Related Work}
\label{sec:related_work}

Automatic database tuning has been studied from several perspectives, including search-based and data-driven tuning, reinforcement-learning-based optimization, language-model-guided methods, and sample-efficient hybrid designs~\cite{dritsas2026mldbms,qiao2025learning,wu2024indexsurvey,siddiqui2023mlindex,chaudhuri1997efficient,ding2020alex,galakatos2019fiting,kraska2018case,liu2020distributed,ma2018query,ma2021mb2,sadri2020online,schnaitter2010semi,tan2019ibtune,van2017automatic,van2021inquiry}.

\textbf{Search-based and data-driven tuning.}
Early automatic tuning methods typically treat the DBMS as a black-box optimization target and search the configuration space through repeated trials~\cite{mahgoub2017rafiki,zhang2021cdbtuneplus,mahgoub2020optimuscloud}. BestConfig~\cite{BestConfig} partitions the configuration space and explores promising regions, while data-driven systems such as OtterTune~\cite{van2017automatic} reuse historical workload--configuration--performance observations to recommend settings for new workloads. E2ETune~\cite{huang2024e2etune} further learns workload-to-configuration mappings with a fine-tuned generative model. These methods show the value of historical observations, but their effectiveness may be limited when prior knowledge is insufficient or when the target deployment differs substantially from previously observed environments.

\textbf{Reinforcement-learning-based tuning.}
Another line of work formulates database tuning as a sequential decision-making problem~\cite{ye2023parameters,shi2023fastune,li2025adwtune,li2024sampleaware,lai2025e2rlixt,wu2022dynamicindex,CDBTune}. CDBTune~\cite{CDBTune} applies DDPG-style reinforcement learning to continuous knob optimization, and Qtune~\cite{li2019qtune} studies query-aware RL-based tuning. HUNTER~\cite{HUNTER} combines warm-start exploration, Random-Forest-based knob ranking, PCA-based state compression, and deep reinforcement learning. These studies demonstrate the potential of RL-based tuning, but also expose practical challenges such as cold-start cost, sample inefficiency, and deployment overhead under limited tuning budgets.

\textbf{Language-model-guided tuning.}
Recent studies use manuals, blogs, forums, and other textual sources to guide database tuning with language models~\cite{dou2025demotuner,zhao2025lgtune,zhang2026sysinsight,li2024llmknob,DB-BERT,GPTuner}. DB-BERT~\cite{DB-BERT} extracts knob-specific hints and recommended values from text corpora for hint-guided optimization, while GPTuner~\cite{GPTuner} integrates heterogeneous knowledge sources, workload-aware knob selection, and Coarse-to-Fine Bayesian optimization. Other methods such as \(\lambda\)-Tune~\cite{lambdatune} and LaTuner~\cite{Latuner} further explore LLM-based tuning strategies. These methods can provide informative initialization, but textual guidance alone may not fully exploit the remaining optimization room of the target system.

\textbf{Sample-efficient and hybrid tuning.}
Recent work also improves tuning efficiency by reducing or reshaping the search space~\cite{seo2023dark,lee2024k2vtune,chai2022cxtuning,cao2024etune,kanellis2022llamatune,HUNTER}. LlamaTune~\cite{kanellis2022llamatune} improves sample efficiency through search-space shaping, dimensionality reduction, and biased handling of special values, while HUNTER~\cite{HUNTER} combines warm-start exploration, reduced-space analysis, and RL-based refinement. These studies suggest that search-space reduction and staged optimization are important for effective tuning under limited budgets.

Prior work offers two complementary insights for improving database tuning efficiency. Reduced-space tuners such as HUNTER show that focusing reinforcement learning on important knobs and compact states can make later-stage exploration more efficient under limited budgets~\cite{HUNTER}. Text-guided tuners such as DB-BERT and GPTuner show that documentation-derived knowledge can quickly move the DBMS toward promising configurations~\cite{DB-BERT,GPTuner}. These insights are useful at different stages of the tuning process: textual guidance is especially helpful for early-stage initialization, while reduced-space reinforcement learning is more suitable for continued refinement after a strong region has been reached. PLRTune therefore organizes these ideas into a staged pipeline. It first constructs workload-specific reduced action and state spaces, then refines textual hints through execution feedback to obtain a stronger initialization, and finally continues state-aware TD3 refinement within the reduced subspace. This design allows prior knowledge to accelerate early progress while preserving execution-driven optimization for the remaining tuning room.

\section{Conclusion}
\label{sec:conclusion}

This paper presents PLRTune, a staged automatic database tuning system built on importance pre-sampling and reranking, execution-guided hint refinement, and state-aware post-tuning refinement. Rather than treating all knobs uniformly from the beginning, PLRTune first identifies workload-relevant important knobs and derives a compact state representation, then refines textual hints to construct a stronger and more reliable initialization, and finally performs TD3-based refinement within the reduced knob subspace to further improve final tuning quality.

Extensive experiments on MySQL and PostgreSQL show that this staged design is effective across diverse workloads. PLRTune achieves the best final result on all nine tested workloads under the same end-to-end evaluation setting. Compared with the corresponding best baselines, PLRTune improves the final objective by \textbf{9.50\%} on average across all workloads, and by \textbf{10.67\%} on average over the eight workloads evaluated by TPS/p95. Runtime-efficiency analysis further shows that PLRTune reaches the strongest baseline's best performance level \textbf{9.03$\times$ faster} on average across workloads. Under the stricter own-best criterion, PLRTune usually reaches its own stronger best result faster than the strongest baseline reaches its own best result, with only two workload-level exceptions where additional time is spent improving beyond the baseline target. Ablation studies confirm that workload-aware search-space reduction, execution-guided hint refinement, and state-aware TD3-based post-tuning refinement all contribute meaningfully to the final result. Cross-hardware experiments further show that workload-specific important-knob priors can remain useful across differently configured target servers when followed by local initialization and refinement.

Overall, the results suggest that combining workload-aware prior construction, stronger hint-guided initialization, and state-aware reduced-space refinement provides an effective way to achieve both fast tuning progress and strong final tuning performance under practical end-to-end tuning budgets.







\section*{Acknowledgment}

The work was supported by the Strategic Priority Research Program of the Chinese Academy of Sciences under Grant No.~XDA0360202.

\end{document}